\documentclass[12pt]{article}

\usepackage{amsmath}
\usepackage{amssymb}
\usepackage{epsfig}
\usepackage{graphicx}
\usepackage{cite}
\usepackage{multirow}
\usepackage{longtable}
\usepackage{lscape}
\usepackage{bbm}

\allowdisplaybreaks[4] 

\newcommand{\capdef}{}
\newcommand{\mycaption}[2][\capdef]{\renewcommand{\capdef}{#2}%
        \caption[#1]{{\footnotesize #2}}}
\makeatletter
\renewcommand{\fnum@table}{\textbf{\tablename~\thetable}}
\renewcommand{\fnum@figure}{\textbf{\figurename~\thefigure}}
\makeatother

\newcounter{myenumi}

\renewcommand{\themyenumi}{\roman{myenumi}}
{\end{list}}

\newlength{\myem}
\newcounter{mysubequation}[equation]

\makeatletter
\renewcommand{\section}{\@startsection{section}{1}{0em}{-\baselineskip}%
{\baselineskip}{\normalfont\large\bfseries}}
\renewcommand{\subsection}%
{\@startsection{subsection}{2}{0em}{-0.7\baselineskip}%
{0.7\baselineskip}{\normalfont\bfseries}}
\makeatother

\newcommand{\bi}{\begin{itemize}}
\newcommand{\ei}{\end{itemize}}

\newcommand{\be}{\begin{equation}}
\newcommand{\ee}{\end{equation}}
\newcommand{\bea}{\begin{eqnarray}}
\newcommand{\eea}{\end{eqnarray}}

\newcommand{\ldm}{\Delta m_{31}^2}
\newcommand{\sdm}{\Delta m_{21}^2}
\newcommand{\deltacp}{\delta_{\mathrm{CP}}}
\newcommand{\stheta}{\sin^2 2 \theta_{13}}
\newcommand{\deltacpt}{\delta_{\mathrm{CP}}^\mathrm{true}}
\newcommand{\sthetat}{\sin^2 2 \theta_{13}^\mathrm{true}}

\newcommand{\ie}{{\it i.e.}}

\newcommand{\eg}{{\it e.g.}}

\newcommand{\cf}{{\it cf.}}

\newcommand{\eq}{Eq.}
\newcommand{\eqs}{Eqs.}

\newcommand{\fig}{Fig.}

\newcommand{\Ref}{Ref.}
\newcommand{\Refs}{Refs.}
\newcommand{\Sec}{Sec.}

\newcommand{\Tab}{Table}

\newcommand{\eet}{\epsilon^m_{e\tau}}
\newcommand{\emt}{\epsilon^m_{\mu\tau}}
\newcommand{\ett}{\epsilon^m_{\tau\tau}}
\newcommand{\eee}{\epsilon^m_{ee}}
\newcommand{\eem}{\epsilon^m_{e\mu}}
\newcommand{\emm}{\epsilon^m_{\mu\mu}}
\newcommand{\eeta}{|\epsilon^m_{e\tau}|}
\newcommand{\emta}{|\epsilon^m_{\mu\tau}|}
\newcommand{\etta}{|\epsilon^m_{\tau\tau}|}
\newcommand{\eeea}{|\epsilon^m_{ee}|}
\newcommand{\eema}{|\epsilon^m_{e\mu}|}
\newcommand{\emma}{|\epsilon^m_{\mu\mu}|}
\newcommand{\eetp}{\phi^m_{e\tau}}
\newcommand{\emtp}{\phi^m_{\mu\tau}}

\newcommand{\equ}[1]{\eq~(\ref{equ:#1})}
\newcommand{\figu}[1]{\fig~\ref{fig:#1}}

\newcommand{\as}[2]{\left( \begin{array}{c} #1 \\ #2 \end{array} \right)}

\begin{document}

\begin{titlepage}

\renewcommand{\thefootnote}{\alph{footnote}}

\vspace*{-3.cm}
\begin{flushright}
\end{flushright}


\renewcommand{\thefootnote}{\fnsymbol{footnote}}
\setcounter{footnote}{-1}

{\begin{center}
{\large\bf
Neutrino factory optimization for non-standard interactions
} \end{center}}
\renewcommand{\thefootnote}{\alph{footnote}}

\vspace*{.8cm}
\vspace*{.3cm}
{\begin{center} {\large{\sc
 		Joachim~Kopp\footnote[1]{\makebox[1.cm]{Email:}
                jkopp@mpi-hd.mpg.de},
 		Toshihiko~Ota\footnote[2]{\makebox[1.cm]{Email:}
                toshihiko.ota@physik.uni-wuerzburg.de}, and
                Walter~Winter\footnote[3]{\makebox[1.cm]{Email:}
                winter@physik.uni-wuerzburg.de}
                }}
\end{center}}
\vspace*{0cm}
{\it
\begin{center}

\footnotemark[1]
	Max-Planck-Institut f{\"u}r Kernphysik, Postfach 103980, D-69029 Heidelberg, Germany

\footnotemark[2]${}^,$\footnotemark[3]
       Institut f{\"u}r Theoretische Physik und Astrophysik, Universit{\"a}t W{\"u}rzburg, \\
       D-97074 W{\"u}rzburg, Germany

\end{center}}

\vspace*{1.5cm}

{\Large \bf
\begin{center} Abstract \end{center}  }

We study the optimization of a neutrino factory with respect to non-standard
neutral current neutrino interactions, and compare the results to those obtained without
non-standard interactions. We discuss the muon energy, baselines, and
oscillation channels as degrees of freedom. Our conclusions are based on
both analytical calculations and on a full numerical simulation of the
neutrino factory setup proposed by the international design study
({\sf IDS-NF}). We consider all possible non-standard parameters,
and include their complex phases. We identify the impact of the
different parameters on the golden, silver, and disappearance
channels. We come to the conclusion that, even in the presence of non-standard
interactions, the performance of the neutrino factory hardly profits from
a silver channel detector, unless the muon energy is significantly increased compared to
the {\sf IDS-NF} setup. Apart from the
dispensable silver channel detector, we demonstrate that the {\sf IDS-NF} setup
is close to optimal even if non-standard interactions are considered.
We find that one very long baseline is a key component in the search for
non-standard interactions, in particular for $\emta$ and $\etta$.

\vspace*{.5cm}

\end{titlepage}

\newpage

\renewcommand{\thefootnote}{\arabic{footnote}}
\setcounter{footnote}{0}

\section{Introduction}

In neutrino physics, three-flavor oscillations have been successfully used as
a model explaining all relevant neutrino data, see, \eg, \Ref~\cite{GonzalezGarcia:2007ib}.
In particular, the solar and atmospheric oscillation parameters have been
measured with very high precisions, and the reactor mixing angle $\theta_{13}$
has been strongly constrained. Future experiments will test this small angle
further, and be sensitive to leptonic CP violation and the neutrino mass
hierarchy (see \Ref~\cite{Bandyopadhyay:2007kx} and references therein). The ultimate
high precision instrument for these purposes might be a neutrino factory~\cite{Geer:1998iz,DeRujula:1998hd,Cervera:2000kp}.
Using different baselines and oscillation channels, it can basically disentangle
all of the remaining oscillation parameters~\cite{Donini:2002rm,Autiero:2003fu,Huber:2003ak,Huber:2006wb} 
in spite of the presence of intrinsic correlations and degeneracies~\cite{Fogli:1996pv,Cervera:2000kp,Minakata:2001qm,Huber:2002mx}.
Because of its high precision, it might be natural to ask how sensitive it is to non-standard
physics.

Non-standard neutrino interactions (NSI) could be such messengers
of new physics beyond the Standard Model in the neutrino sector.
In this work, we will focus on interactions of the form 
\begin{equation}
\left\{
 \overline{\nu}_{\beta} \gamma^{\rho} (1-\gamma^5)
 \nu_{\alpha}
 \right\}
 \left\{
 \bar{f} \gamma_{\rho} (1 \pm\gamma^5) f
 \right\},
\nonumber
\end{equation}
which may affect the neutrino propagation in matter~\cite{Wolfenstein:1977ue,Valle:1987gv,Guzzo:1991hi,Roulet:1991sm}. Here, $f$ is an electron or a first-generation quark. 
Such dimension six operators 
can be considered as an effective low-energy fingerprint
of new physics at a higher energy scale, once the high energy
degrees of freedom have been integrated out.
Note that these operators
involve the same in- and out-state charged fermion, which means
that they produce neutral current-like interactions. Similarly, one can write down operators
involving different in- and out-state charged fermions, leading to non-standard effects in the neutrino production and detection~\cite{Grossman:1995wx}.
From the theory point of view,
dimension six operators suffer from the problem that the
neutrinos come together with their SU(2) counterparts in the Standard
Model, which means that charged lepton flavor-violating (LFV) processes are introduced
at tree level (unless the SU(2) breaking effects are large)\footnote{%
There is a possibility to construct the dimension six operator without SU(2) counter processes,
assuming a charged SU(2) singlet mediation~\cite{Bergmann:1999pk}.
However, it is constrained by the measurement of the Fermi constant and 
the lepton universality~\cite{Bergmann:1999pk,Babu:2002uu}.
}; see, \eg, \Refs~\cite{Bergmann:1999rz,Bergmann:1999pk,Bergmann:1998ft,Berezhiani:2001rs}. 
For example, 
if $\alpha=e$, $\beta=\tau$, and $f=e$, $\tau$ decays into three electrons are a consequence,
which can be strongly constrained by B-factories.
However, this SU(2) relation can be avoided when dimension eight operators 
are taken into account above the electroweak scale~\cite{Berezhiani:2001rs,Davidson:2003ha}.
In such a case, charged LFV effects appear only at one loop level, 
leading to less stringent bounds.
Of course, there are ways to circumvent this reasoning, but so far there
is no motivation to assume that the NSI should be large. Therefore, we focus on
NSI constraints in this study. For a summary of current bounds from non-oscillation experiments,
see \Refs~\cite{Davidson:2003ha,Barranco:2007ej,Bandyopadhyay:2007kx}.

NSI are also constrained from the current oscillation experiments.
The effect of NSI in solar neutrino experiments has been studied in 
\Refs~\cite{Friedland:2004pp,Guzzo:2004ue,Miranda:2004nb}, and in 
atmospheric neutrino oscillation in 
\Refs~\cite{Gonzalez-Garcia:1998hj,Bergmann:1999pk,Fornengo:2001pm,Gonzalez-Garcia:2004wg,Friedland:2004ah,Friedland:2005vy}.
Some numbers on the constraints can be found in these references.
In addition, NSI have been discussed in the context of future neutrino oscillation experiments
in \Refs~\cite{Bueno:2000jy,Huber:2001de,Huber:2001zw,Gonzalez-Garcia:2001mp,Ota:2001pw,Gago:2001xg,Campanelli:2002cc,Ota:2002na,Huber:2002bi,Hattori:2002uw,Garbutt:2003ih,Blennow:2005qj,Friedland:2006pi,Kitazawa:2006iq,Honda:2006gv,Adhikari:2006uj,Blennow:2007pu,Kopp:2007mi,Ribeiro:2007ud,Kopp:2007ne,Ribeiro:2007jq,EstebanPretel:2008qi}. They can be also tested in astrophysical neutrino sources, such as supernovae~\cite{Fogli:2002xj,EstebanPretel:2007yu}, and in the early universe~\cite{Mangano:2006ar}.
The sensitivity of non-oscillation experiments to NSI, such as collider and neutrino scattering experiments, has been pointed out in \Refs~\cite{Berezhiani:2001rs,Chen:2007cn,Barranco:2005ps,Berezhiani:2001rt,Barranco:2005yy,Barranco:2007tz}.

As far as the neutrino factory is concerned, the problem of parameter correlations and degeneracies has to be taken into account if one parameter needs to be extracted from the information encoded in the event rates. NSI might therefore be confused with the standard oscillation parameters, and the sensitivity to the standard oscillation parameters might be affected in the presence of NSI~\cite{Huber:2002bi,Huber:2001de}. The discovery reach to NSI (including production and detection effects) has been studied in \Ref~\cite{Kopp:2007mi} for a setup with only one detector. A major step forward was taken in \Ref~\cite{Ribeiro:2007jq}, where it was pointed out that a detector at the magic baseline has an excellent sensitivity to interactions of the form $\nu_e + f \leftrightarrow \nu_\tau + f$. On the other hand, the silver channel $\nu_e \rightarrow \nu_\tau$~\cite{Donini:2002rm,Autiero:2003fu} is very sensitive to this particular interaction at high energies~\cite{Kitazawa:2006iq}, because it is the leading order effect in that channel~\cite{Bandyopadhyay:2007kx}. Therefore, it is yet unclear what the best strategy to measure specific NSI actually is, and what the role of the silver channel for NSI could be. We will clarify the contribution of different channels to specific NSI in neutrino propagation in this study both analytically and numerically. In addition, we discuss the muon energy and baseline optimization of a neutrino factory for NSI, and compare it to the optimization for the standard oscillation (SO) parameters. We fully take into account complex phases, and we include all possible NSI in the discussion~--~we will comment on specific parameters in the following section. We adopt the point of view that NSI must be small, since there is not yet any theoretical motivation for large NSI. Therefore, we only discuss sensitivity limits, and no discovery reaches. We focus on neutrino propagation effects for the sake of simplicity, which means that we assume that production and detection effects are either not present, or constrained otherwise (such as by a near detector).
Our starting point for the optimization will be the current baseline setup for the international design study of a neutrino factory ({\sf IDS-NF})~\cite{Bandyopadhyay:2007kx,ids}, which is designed for optimal discovery reaches for $\stheta$, the neutrino mass hierarchy, and leptonic CP violation. 

Our study is organized as follows: We discuss all possible NSI in neutrino propagation in \Sec~\ref{sec:pheno}, identify the ones relevant for a neutrino factory, and describe the channels providing the main sensitivities analytically. In \Sec~\ref{sec:performance}, we introduce our performance indicators and simulation details. The impact of different channels on specific types of non-standard parameters is, in a full simulation, illustrated in \Sec~\ref{sec:channels}. We then study the impact of the muon energy in \Sec~\ref{sec:muonen}, where we put special emphasis on the silver channel. The baseline optimization of the two-baseline setup is, for standard oscillation physics, revisited in \Sec~\ref{sec:baselines}, and it is compared to the same optimization for the NSI parameters. In \Sec~\ref{sec:summary}, we summarize the sensitivities expected from a neutrino factory,  and we conclude.

\section{Phenomenology}
\label{sec:pheno}

We focus on non-standard propagation effects
in neutrino oscillations. These can be phenomenologically 
described by neutral current-type non-standard interactions (NSI)  
\begin{eqnarray}
\mathcal{L}_{\rm NSI}
 & =&
 (G_{F} / \sqrt{2})
 (\epsilon^{fP}_{\beta \alpha})
 \left\{
 \overline{\nu}_{\beta} \gamma^{\rho} L
 \nu_{\alpha}
 \right\}
 \left\{
 \bar{f} \gamma_{\rho} P f
 \right\}
 \nonumber \\
 &+&
  (G_{F} / \sqrt{2})
 (\epsilon^{fP}_{\beta \alpha})^{*}
 \left\{
 \overline{\nu}_{\alpha} \gamma^{\rho} L
 \nu_{\beta}
 \right\}
 \left\{
 \bar{f} \gamma_{\rho} P f
 \right\},
\label{equ:lagrangian}
\end{eqnarray}
affecting 
the neutrino propagation in matter. Here,
$\epsilon^{fP}_{\alpha \beta} = (\epsilon^{fP}_{\beta \alpha})^{*}$, $P \in \{L,R\}$, $L=1 - \gamma^5$, $R=1+\gamma^5$, and $f$ stands for all possible fermions in Earth matter ($u$ quarks, $d$ quarks, electrons). This definition includes the possibility of different non-standard
interactions with quarks and leptons, and different interactions for left- and right-handed
couplings to the fermions.\footnote{The coupling to the neutrino fields is left-handed because a right-handed coupling would be either helicity-suppressed, or only present in higher order corrections (there have to be at least two non-standard vertices in the amplitude to produce and absorb the right-handed neutrino).}
 Note that, in general, $\epsilon^{fP}_{\alpha \beta}$ are complex numbers for
$\alpha \neq \beta$, and real numbers for $\alpha = \beta$.
Since there are about two nucleons (a proton and a neutron) per electron in Earth matter, 
neutrinos are, for coherent forward scattering in Earth matter, sensitive to
the combination 
\begin{equation}
\epsilon^{m}_{\beta \alpha} = 3 \epsilon^{u}_{\beta \alpha}+
3 \epsilon^{d}_{\beta \alpha} + \epsilon^{e}_{\beta \alpha} \, , 
\end{equation}
where $\epsilon^{f}_{\beta \alpha} \equiv \epsilon^{fL}_{\beta \alpha}+\epsilon^{fR}_{\beta \alpha}$. This is because the neutrino beams are only sensitive to the vector component.
Further on, we will discuss how well one can test this combination. For the bounds
on interactions for individual fermions, see \Tab~8 in \Ref~\cite{Bandyopadhyay:2007kx}. 

\subsection{Neutrino propagation Hamiltonian and considered NSI parameters}
\label{sec:prop}

Interactions of the type in \equ{lagrangian}
add an extra effective matter effect potential
to the neutrino propagation Hamiltonian, which then
reads
\begin{eqnarray}
 H & = &
 \frac{1}{2E}
 \left\{
 U
 \begin{pmatrix}
  0 && \\
  & \Delta m_{21}^{2} & \\
  && \Delta m_{31}^{2}
 \end{pmatrix}
 U^{\dagger}
 +
 \begin{pmatrix}
  a_{\rm CC} && \\
  & 0 & \\
  && 0
 \end{pmatrix}
 + \right. \nonumber \\
& + &  
\left.  a_{\rm CC}
 \begin{pmatrix}
  \epsilon^{m}_{ee} &  \epsilon^{m}_{e\mu} &  \epsilon^{m}_{e\tau} \\
   (\epsilon^{m}_{e\mu})^{*}  &  \epsilon^{m}_{\mu\mu} 
  &  \epsilon^{m}_{\mu\tau} \\
   (\epsilon^{m}_{e\tau})^{*} &  (\epsilon^{m}_{\mu\tau})^{*} &  
  \epsilon^{m}_{\tau\tau}
 \end{pmatrix}
 \right\}.
\label{equ:Ham}
\end{eqnarray}
Here, $a_{\rm CC}$ is the usual matter effect term defined 
as $a_{\rm CC} \equiv 2 \sqrt{2} E G_{F} N_{e}$ (with $N_e$ the electron number density in
Earth matter), and the first line
corresponds to the usual Hamiltonian in Earth matter. This equation already
implies that the energy and baseline dependence of the non-standard effects  
will be similar to the standard matter effects, \ie, long baselines and
high neutrino energies are important. For antineutrinos, the matter potential in \equ{Ham}
and all complex phases change sign, \ie,  $a_{\rm CC} \rightarrow -a_{\rm CC}$, $U \rightarrow U^*$, and
$\epsilon^m_{\beta \alpha} \rightarrow (\epsilon^m_{\beta \alpha})^*$.
Note that from the hermiticity of the Hamiltonian, $\epsilon_{\alpha \alpha}^m$ are
real numbers, while the $\epsilon_{\alpha \beta}^m$'s can be complex for $\alpha \neq \beta$.
As far as purely phenomenological bounds are concerned, $\eema$  and $\emma$
are already very well constrained (see, \eg, \Tab~8 in \Ref~\cite{Bandyopadhyay:2007kx}).
In fact, we will show at the end of this study, that the bounds obtainable from the neutrino factory
are comparable to the current bounds, which means that the neutrino factory is probably not the
best experiment for their measurement.
The interaction described by $\eee$ is not {\em per se} interesting for us, since it will be intimately
correlated with the matter density. We will discuss it in \Sec~\ref{sec:muonen}.

Because of the strong bounds on $\eema$  and $\emma$, and the straightforward relationship between $\eee$ and the matter density precision measurement, we will focus on $\eet$, $\emt$, and $\ett$ in the main line of this study. 
Note that the above mentioned bounds are purely phenomenological, and there
are no convincing theoretical arguments yet why these non-standard effects should be large.
Hence we focus on further constraints beyond the current limits in this study,
but we do not discuss a possible discovery of non-standard effects, and only marginally touch possible effects on the
determination of the standard oscillation parameters.  
Note that similar non-standard effects can be present in the neutrino production or detection. 
We do not consider these effects, which has the advantage that we do not have to simulate the near detector explicitely.\footnote{There is not yet any near detector specification in the {\sf IDS-NF} baseline setup. As soon as such a specification is available, it may make sense to discuss production and detection effects as well.}

\subsection{Measuring $\boldsymbol{\eet}$ in the golden and silver appearance channels}

Let us now first of all focus on $\eet$, which can be best measured in the golden $\nu_e \rightarrow \nu_\mu$ and silver
$\nu_e \rightarrow \nu_\tau$ appearance channels (see, \eg, \Ref~\cite{Kitazawa:2006iq}). 
The interference term induced by  $\eet$ in the silver channel $P_{e \tau} \equiv P_{\nu_{e} \rightarrow \nu_{\tau}} $ can be illustrated as
\begin{equation}
       P_{e \tau}
       =
       \underbrace{
       \left|
       \mathcal{A} (\nu_{e} \xrightarrow[]{\text{SO}} \nu_{\tau})
       \right|^{2}
       }_{
       \begin{minipage}{2cm}
	\begin{center}
	\tiny
       SO signal, \\
	background for NSI search
	\end{center}
       \end{minipage}}
       +
       \underbrace{
       2 {\rm Re} \left[
       \mathcal{A}^{*} (\nu_{e} \xrightarrow[]{\text{SO}} \nu_{\tau})
       \mathcal{A} (\nu_{e} \xrightarrow[\epsilon^{m}_{e\tau}]
       {\text{NSI, No-osc}} \nu_{\tau})
       \right]
       +\mathcal{O}(\eeta^{2})
       }_{\text{NSI signal}} \, ,
\label{equ:schematic}
\end{equation}
where ``SO'' stands for ``standard oscillations''. As we will see below, this structure
is recovered in the full expression of the oscillation probability. In our discussion, we
will use the following abbreviations for the spectral terms, \ie, the terms containing energy and/or
baseline information:
\bea
\label{equ:s1}
\Delta & \equiv & \frac{\ldm L}{4 E} \, , \\
\hat{A} & \equiv & \pm \frac{a_{\mathrm{CC}}}{\ldm} = \pm \frac{2 \sqrt{2} E G_F N_e}{\ldm} , \\
\mathcal{F}^{\mathrm{Res}} & \equiv & \frac{\sin[ ( 1 - \hat{A}) \Delta]}{1-\hat{A}} \, , \\
\mathcal{F}^{\mathrm{MB}} & \equiv & \sin( \hat{A} \Delta ) = \sin( \pm \frac{\sqrt{2}}{2} G_F N_e L ) \, .
\label{equ:s4}
\eea
Here $\Delta$ corresponds to the vacuum oscillation phase, $\hat{A}$ to the effective matter potential with $\hat{A} \rightarrow 1$ at the matter resonance, $\mathcal{F}^{\mathrm{Res}}$ to a term maximal at the matter resonance, and $\mathcal{F}^{\mathrm{MB}}$ to a term which is vanishing at the magic baseline $L \simeq \mathrm{7 \, 500 \, km}$~\cite{Huber:2003ak,Smirnov:2006sm}. In the definitions of $\hat{A}$ and $\mathcal{F}^{\mathrm{MB}}$, the upper signs are for neutrinos, and the lower ones for antineutrinos.
Contributions proportional to different products of these terms can, in principle, be disentangled by the use of a wide beam spectrum and different baselines.
The standard oscillation probability
for $P_{e \mu}$ and $P_{e \tau}$ is, to second order in $\alpha \equiv \sdm/\ldm \simeq 0.03$ and $\sin 2 \theta_{13}$, given by (see, \eg, \Ref~\cite{Akhmedov:2004ny})
\begin{eqnarray}
\as{P_{e \mu}^{\mathrm{SO}}}{P_{e \tau}^{\mathrm{SO}}} & \simeq & \stheta \as{s_{23}^2}{c_{23}^2} (\mathcal{F}^{\mathrm{Res}})^2 \nonumber \\
& \pm &  \alpha \, \sin 2 \theta_{13} \, \sin 2 \theta_{12} \, \sin 2 \theta_{23} \, \sin \deltacp  \, \frac{1}{\hat{A}} \, \mathcal{F}^{\mathrm{MB}} \, \mathcal{F}^{\mathrm{Res}} \, \sin \Delta \nonumber \\
& \pm &  \alpha \, \sin 2 \theta_{13} \, \sin 2 \theta_{12} \, \sin 2 \theta_{23} \, \cos \deltacp \, \frac{1}{\hat{A}} \, \mathcal{F}^{\mathrm{MB}} \, \mathcal{F}^{\mathrm{Res}} \,  \cos \Delta \nonumber \\
& + & \alpha^2 \as{c_{23}^2}{s_{23}^2} \, \sin^2 2 \theta_{12} \, \frac{1}{\hat{A}^2} \, (\mathcal{F}^{\mathrm{MB}})^2 \, 
\label{equ:so}
\end{eqnarray}
with $s_{ij}=\sin \theta_{ij}$ and $c_{ij}=\cos \theta_{ij}$. Note that the upper row/signs are for $P_{e \mu}$, and the lower row/signs for  $P_{e \tau}$. The different terms, can in principle, be disentangled by their spectral dependencies. For example, for $\mathcal{F}^{\mathrm{MB}} \rightarrow 0$ (magic baseline) only the first term survives, which allows for a clean measurement of $\stheta$ and the mass hierarchy. The relative amplitude of the different terms is given by the size of $\stheta$ compared to $\alpha^2 \simeq 0.001$: For $\stheta \gg \alpha^2$, the first term dominates, for $\stheta \simeq \alpha^2$, all terms including the middle (CP-terms) are large, and for $\stheta \ll \alpha^2$, the last (solar) term dominates.
Introducing non-standard effects by $\eet$, we have to second order in $\alpha$, $\sin 2 \theta_{13}$, and $\eeta$
\begin{eqnarray}
\as{P_{e \mu}^{\mathrm{NSI}}}{P_{e \tau}^{\mathrm{NSI}}}& \simeq & \as{P_{e \mu}^{\mathrm{SO}}}{P_{e \tau}^{\mathrm{SO}}}  \nonumber \\
& \mp & 2 \, \eeta \, \sin 2 \theta_{13} \, \sin 2 \theta_{23} \, s_{23} \, \sin( \deltacp + \eetp) \, \mathcal{F}^{\mathrm{MB}} \, \mathcal{F}^{\mathrm{Res}} \,  \sin \Delta \nonumber \\
& \mp & 2 \, \eeta \, \sin 2 \theta_{13} \, \sin 2 \theta_{23} \, s_{23} \, \cos( \deltacp + \eetp) \, \mathcal{F}^{\mathrm{MB}} \, \mathcal{F}^{\mathrm{Res}} \,  \cos \Delta \nonumber \\
& + & 4 \, \eeta \, \sin 2 \theta_{13} \, c_{23} \, \as{s_{23}^2}{c_{23}^2} \, \cos( \deltacp + \eetp) \, \hat{A} \, (\mathcal{F}^{\mathrm{Res}})^2 \nonumber \\
&\mp & 2 \, \eeta \, \alpha \, \sin 2 \theta_{12} \, \sin 2 \theta_{23} \, c_{23} \, \sin \eetp \, \mathcal{F}^{\mathrm{MB}} \, \mathcal{F}^{\mathrm{Res}} \,  \sin \Delta \nonumber \\ 
&\pm & 2 \, \eeta \, \alpha \, \sin 2 \theta_{12} \, \sin 2 \theta_{23} \, c_{23} \, \cos \eetp \, \mathcal{F}^{\mathrm{MB}} \, \mathcal{F}^{\mathrm{Res}} \,  \cos \Delta \nonumber \\  
& - & 4 \, \eeta \, \alpha \, \sin 2 \theta_{12} \, s_{23} \, \as{c_{23}^2}{s_{23}^2} \, \cos \eetp \, \frac{1}{\hat{A}} \, (\mathcal{F}^{\mathrm{MB}})^2 \nonumber \\
& + & 4\, \eeta^2 \, c_{23}^2 \, \as{s_{23}^2}{c_{23}^2} \, \hat{A}^2 \, (\mathcal{F}^{\mathrm{Res}})^2  \nonumber \\
& \mp & 2 \, \eeta^2 \, \sin^2 2 \theta_{23} \, \hat{A} \, \mathcal{F}^{\mathrm{MB}} \, \mathcal{F}^{\mathrm{Res}} \, \cos \Delta \nonumber \\
& + & 4 \, \eeta^2 \, s_{23}^2 \as{c_{23}^2}{s_{23}^2} \,  (\mathcal{F}^{\mathrm{MB}})^2 \, .
\label{equ:nsi}
\end{eqnarray}
For antineutrinos, $\hat{A}$ changes sign, and all phases are inverted, \ie, the corresponding $\sin$-terms change signs. That means that the 2nd and 5th terms in \equ{nsi} are the CP-odd terms describing intrinsic non-standard CP violation. Note that there can be CP violation even for $\stheta=0$, which is then induced by the 5th term.
As we can read off from \equ{schematic}, there are terms proportional to $\eeta \, \sin 2 \theta_{13}$, terms proportional to $\eeta \, \alpha$, and terms proportional to $\eeta^2$, which dominate depending on the relative size of $\sin 2 \theta_{13}$, $\alpha$, and $\eeta$. For example, for $\stheta=0$, only the last six terms survive. If in addition $\eeta \gg \alpha \simeq 0.03$, the last three terms dominate, which are quadratic in $\eeta$.

 For the current best-fit value $\theta_{23}=\pi/4$, $P_{e \mu}^{\mathrm{NSI}}$ and $P_{e \tau}^{\mathrm{NSI}}$ differ only by the signs as given in the 2nd, 3rd, 5th, 6th, and 9th terms (and the 2nd and 3rd term in \equ{so}). If there is information from many different baselines and energies, the different dependencies on the spectral terms 
\eqs~(\ref{equ:s1}) to~(\ref{equ:s4}) can be used to disentangle all terms in \equ{nsi} except for the 2nd and 5th (or 3rd and 6th) terms. Depending on the relative size of $\sin 2 \theta_{13}$ and $\alpha$, either of these two terms may dominate, or both terms might be of similar magnitude. Note, however, that in certain limits, \equ{nsi} is very different for the golden and silver channels.
For example, let us consider the situation at peak energies of the
spectrum, and at the first oscillation maximum, which occurs
typically at a baseline around $L \sim 3000$ to $4000$~km.
In this case, $\hat{A} \Delta \simeq \pi/2$ and $\Delta \ll 1$.
It is easy to see from \equ{s1} to \equ{s4}, and from \equ{so}
that this leads to all standard oscillation terms being
$\propto \Delta^2 \propto 1/E^2$, so that they cannot be
disentangled from each other. Of the NSI terms in \equ{nsi},
only those which are either constant in energy, or proportional
to $1/E$, may be separated from the standard terms. From the
signs in \equ{nsi}, we find that, for the golden channel, all
relevant NSI terms cancel, while for the silver channel, they
interfere constructively. Consequently, the golden channel
detector at the short neutrino factory baseline will not be
able to provide a good sensitivity to $\eet$, while a better
performance is expected for a silver channel detector at the
same baseline. Our argument also shows that a high neutrino
energy is advantageous for a measurement of NSI at
$\hat{A} \Delta \simeq \pi/2$ and $\Delta \ll 1$, because
it reduces the standard oscillation background.

At the magic baseline, $\mathcal{F}^{\mathrm{MB}} \rightarrow 0$~\cite{Huber:2003ak}, 
we obtain
\begin{eqnarray}
\as{P_{e \mu}^{\mathrm{NSI}}}{P_{e \tau}^{\mathrm{NSI}}}_{\mathrm{MB}} & \simeq & 
\stheta \as{s_{23}^2}{c_{23}^2} (\mathcal{F}^{\mathrm{Res}}|_{\rm MB})^2 \nonumber \\
& + & 4 \, \eeta \, \sin 2 \theta_{13} \, c_{23} \, \as{s_{23}^2}{c_{23}^2} \, \cos( \deltacp + \eetp) \, \hat{A} \, (\mathcal{F}^{\mathrm{Res}}|_{\rm MB})^2 \nonumber \\
& + & 4\, \eeta^2 \, c_{23}^2 \, \as{s_{23}^2}{c_{23}^2} \, \hat{A}^2 \, (\mathcal{F}^{\mathrm{Res}}|_{\rm MB})^2  \, .
\label{equ:nsimb}
\end{eqnarray}
Here $\mathcal{F}^{\rm Res}|_{\rm MB}$ is $\mathcal{F}^{\rm Res}$ in the 
magic baseline limit $\Delta \hat{A} \rightarrow \pi$. 
This formula is exactly the same as in \Ref~\cite{Ribeiro:2007ud} for $P_{e \mu}$ if $\mathcal{F}^\mathrm{Res}|_{\rm MB}$ is trigonometrically expanded.\footnote{We keep, however, $\mathcal{F}^\mathrm{Res}|_{\rm MB}$ in the formula, because we can even correctly reproduce the resonance limit $\hat{A} \rightarrow 1$. In \Ref~\cite{Ribeiro:2007ud}, $P_{e \mu} \rightarrow \infty$ for $\hat{A} \rightarrow 1$.}
It has a number of interesting implications. First of all, there are much less correlations than in \equ{nsi}, which means that the magic baseline will crucially contribute to the NSI sensitivity. However, compared to the SO case, the NSI case is not completely correlation-free at the magic baseline~--~even the phase $\eetp$ appears in the formula.
Second, for maximal atmospheric mixing, we have $P_{e \mu} = P_{e \tau}$, which means that there is no difference between the golden and silver channels. Therefore, there will be no physics case for the silver channel at the magic baseline because of the much lower event rate.  And third, since all terms are proportional to $(\mathcal{F}^{\mathrm{Res}})^2$, the second term, which is proportional to $\hat{A} \propto E$, and the third term, which is proportional to $\hat{A}^2 \propto E^2$, become relatively enhanced for high energies.
This means that high neutrino energies are very important to constrain NSI. From \equ{nsimb}, we can already estimate that the $\eeta^2$ sensitivity should quantitatively be comparable to the $\stheta$ sensitivity, \ie, if the $\stheta$ sensitivity is about $10^{-5}$, we obtain a 
$\eeta$ sensitivity of about $0.003$ if correlations and degeneracies can be sufficiently resolved. 

\subsection{Measuring $\boldsymbol{\ett}$ and $\boldsymbol{\emt}$ in the disappearance channel}
\label{sec:ettemt}

As we shall quantitatively discuss later, the disappearance channel $P_{\mu \mu}$ at the neutrino factory is the dominant source for the $\emt$ and $\ett$ sensitivities (see, \eg, \Ref~\cite{Honda:2006gv}).
Here we follow \Ref~\cite{Blennow:2005qj} to describe these effects
in the two flavor limit. 
The approximation corresponds to the $\nu_{\mu}$-$\nu_{\tau}$ system 
with $\theta_{13} \rightarrow 0$. For $\ett$, we have 
\begin{equation}
H
 =
 \frac{1}{2 E}
 \left\{
 U
 \begin{pmatrix}
  0 & \\
  & \Delta m_{31}^{2}
 \end{pmatrix}
 U^\dagger
 +
 \begin{pmatrix}
  -a_{\mathrm{CC}} \epsilon^{m}_{\tau\tau} & \\
  & 0
 \end{pmatrix}
 \right\}
 +
 \frac{a_{\mathrm{CC}} \epsilon^{m}_{\tau\tau}}{2E} 
 {\bf 1} \, ,
\end{equation}
where the PMNS matrix $U$ is a $2 \times 2$ mixing matrix with 
the mixing angle corresponding to $\theta_{23}$.
From this expression, we can read off the fact that 
$\ett$ plays the same roll as $\emm$ does~\cite{Honda:2006gv}.
In this case, we can describe the shift in the mass squared difference
and mixing angle by a parameter mapping:
\begin{eqnarray}
\Delta \tilde{m}_{31}^{2}
 & =&
 \Delta m_{31}^{2}
 \sqrt{
 \sin^{2} 2 \theta_{23}
 +
 \left(
 \hat{A} \ett
 + \cos 2\theta_{23}
 \right)^{2}
 }, \\
\sin^{2} 2\tilde{\theta}_{23}
 &=&
 \frac{\sin^{2} 2\theta_{23}}
 {\sin^{2} 2\theta_{23} 
 + 
 \left(
 \hat{A} \ett 
 + \cos 2 \theta_{23} \right)^{2}
 }
\label{equ:t23res}
\end{eqnarray}
In the maximal mixing limit $\theta_{23} \rightarrow \pi/4$, 
they are reduced to
\begin{eqnarray}
 \Delta \tilde{m}_{31}^{2}
 &\rightarrow&
 \Delta m_{31}^{2}
 \sqrt{
 1
 +
 \left(\hat{A} \epsilon^{m}_{\tau\tau}
 \right)^{2}
 }, \label{equ:ldmett} \\
\sin^{2} 2\tilde{\theta}_{23}
 & =&
 \frac{1}
 {1
 + 
 \left(
  \hat{A}  \epsilon^{m}_{\tau\tau}
 \right)^{2}
 }.
\end{eqnarray}
The lowest order of this shift comes from 
$\mathcal{O}\{(\epsilon^{m}_{\tau\tau})^{2}\}$, 
which means that it does not appear in the analytic expressions 
in \Refs~\cite{Kopp:2007ne,Ribeiro:2007jq}.
The NSI effect is proportional to $\hat{A}^2 \propto E^{2}$.
In low energy experiments such as T2K, this effect is not
important. On the other hand, 
in high energy experiments, such as neutrino factories,
this will affect the oscillation probability significantly. In addition, note that there
can, in principle, be resonant effects for strong deviations from maximal mixings.
For $\etta=\mathcal{O}(1)$ (which might be, however, unrealistically large~\cite{Fornengo:2001pm}) and $\theta_{23}$ on the edge of the current $3\sigma$ allowed range, one finds from
\equ{t23res} that
one can have resonance energies as high as about $2.5 \, \mathrm{GeV}$, which is slightly above the currently considered detection threshold. 

For $\epsilon^{m}_{\mu\tau}$, which is also strongly present
in the disappearance channel, the parameter mapping is
slightly more complicated because $\epsilon^{m}_{\mu\tau}$
can have a complex phase $\emtp$:
\begin{eqnarray}
 \Delta \tilde{m}_{31}^{2}
 &=&
 \Delta m_{31}^{2}
 \sqrt{
 \left(2 \hat{A} |\epsilon^{m}_{\mu\tau}|
 \cos {\phi^{m}_{\mu\tau}}
 +
 \sin 2 \theta_{23}
 \right)^{2}
 +
 \left(
 2 \hat{A} |\epsilon^{m}_{\mu\tau}|
 \sin {\phi^{m}_{\mu\tau}}
 \right)^{2}
 +
 \cos^{2} 2 \theta_{23}
 }, \\
\sin^{2} 2 \tilde{\theta}_{23}
 &=&
\frac{
 (
 2 \hat{A}
 |\epsilon^{m}_{\mu\tau}| \cos \phi^{m}_{\mu\tau}
 +
 \sin 2 \theta_{23}
 )^{2}
 +
 \left(
 2 \hat{A}  |\epsilon^{m}_{\mu\tau}| \sin \phi^{m}_{\mu\tau}
 \right)^{2}
 }{
 \left(
 2 \hat{A} |\epsilon^{m}_{\mu\tau}|
 \cos {\phi^{m}_{\mu\tau}}
 +
 \sin 2 \theta_{23}
 \right)^{2}
 +
 \left(
 2 \hat{A} |\epsilon^{m}_{\mu\tau}|
 \sin {\phi^{m}_{\mu\tau}}
 \right)^{2}
 +
 \cos^{2} 2 \theta_{23}
}.
\end{eqnarray}
For maximal mixing $\theta_{23} \rightarrow \pi/4$, we obtain $\sin^{2} 2 \tilde{\theta}_{23} \rightarrow 1$ and 
\begin{equation}
\Delta \tilde{m}_{31}^{2}
\rightarrow
\Delta m_{31}^{2}
 \sqrt{
 1 + 4 \hat{A} 
 |\epsilon^{m}_{\mu\tau}|
 \cos \phi^{m}_{\mu\tau}
 +
 \left(
 2 \hat{A} |\epsilon^{m}_{\mu\tau}|
 \right)^{2}
 } \, .
\label{equ:ldmshift-eMmutau}
\end{equation}
We can see that the mass squared difference receives modifications already
at first order in $\epsilon^{m}_{\mu\tau}$, while the mixing angle remains
maximal to all orders.
Since we will marginalize over the phase of the NSI parameter,
the visible effect comes from the second order 
term. 

Numerically, the sensitivity to $\emta$ and $\etta$ will be limited by the
precision of $\ldm$ (provided that all other correlations can be resolved).
In \Ref~\cite{Huber:2006wb}, the $1\sigma$ precision of $\ldm$ has been
found to be $0.2\%$ at the very long baseline. \equ{ldmett} thus implies
$0.5 \, (\hat{A} \etta)^2 \simeq 0.2\%$ at the $1\sigma$ sensitivity limit
for $\etta$. At the upper end of the neutrino spectrum
($E_\nu \simeq E_\mu = 25 \, \mathrm{GeV}$, $\hat{A} \simeq 3$), this leads
to $\etta \simeq 0.02$.
 From \equ{ldmshift-eMmutau}, we obtain a much better sensitivity for real $\emt$, \ie, $\cos \emtp=\pm 1$: In this case, the sensitivity is linear in $\emta$, and given by $2 \hat{A} \emta \simeq 0.2\%$ at the sensitivity limit, or $\emta \simeq 3 \, \cdot 10^{-4}$.
If, however, $\emtp$ can take any value, it can also assume $\emtp=\pm \pi/2$, and we are back in the quadratic regime such as for $\etta$. In fact, one can even have cancellation of the two terms in \equ{ldmshift-eMmutau}, which means that we expect a sensitivity worse than for $\etta$. 

\section{Performance indicators and simulation details}
\label{sec:performance}

In the previous section, we have motivated why we only consider small
non-standard effects.
As performance indicator, we use the ``$|\epsilon_{\alpha \beta}^m|$ sensitivity'', which
corresponds to the exclusion limit which is obtained if the true value (simulated value) vanishes.
In principle, we follow the same definition as for the $\stheta$ sensitivity, \ie,
we define the $|\epsilon^m_{\alpha \beta}|$ sensitivity as the largest fit $|\epsilon^m_{\alpha \beta}|$
which fits the true $|\epsilon^m_{\alpha \beta}|=0$. Note that $\epsilon^m_{\alpha \beta}$
can be complex for $\alpha\neq\beta$, which means that the (fit) phase $\phi^m_{\alpha \beta}$
has to be marginalized over, whereas the true phase is irrelevant because of the true 
 $|\epsilon^m_{\alpha \beta}|=0$. In addition, all standard oscillation parameters are marginalized over.
In our simulations, we find the main correlation leading to technical difficulties is
the correlation among $\phi^m_{\alpha \beta}$, $\deltacp$, and $\stheta$. Therefore, we pre-scan this
set of parameters in many cases to find the position of the global minimum.
In some cases, we will consider also correlations among different
$\epsilon_{\alpha \beta}^m$'s in order to compare our results to earlier works.
For the same reason, we will sometimes also neglect the phases
$\phi^m_{\alpha \beta}$ even for $\alpha \neq \beta$.
For the sake of simplicity, we do not include the $\mathrm{sgn}(\ldm)$ degeneracy for the non-standard sensitivities~\cite{Minakata:2001qm}. In addition, we do not consider degeneracies with unrealistically large $|\epsilon_{\alpha \beta}^m| \gtrsim 1$ in some cases, because these degeneracies would appear above the current bounds.
Note that our $|\epsilon^m_{\alpha \beta}|$ sensitivity is expected to be similar to a conservative case discovery limit, \ie, depending on the phases, the discovery may be possible for smaller $|\epsilon^m_{\alpha \beta}|$ than the sensitivity limit (\cf, \Refs~\cite{Kopp:2007ne,Kopp:2007mi}). 

The experimental scenario we consider is the {\sf IDS-NF}~1.0 setup from \Ref~\cite{ids}, which is the current standard setup for the ``International design study of the neutrino factory'' ({\sf IDS-NF}). Within the ``International scoping study of a future neutrino factory and super-beam facility''~\cite{Huber:2006wb,Bandyopadhyay:2007kx}, this setup has been optimized for the measurement of $\stheta$, the neutrino mass hierarchy, and leptonic CP violation in the case of standard oscillations.
In short, this setup uses two baselines at about $4 \, 000 \, \mathrm{km}$ and $7 \, 500 \, \mathrm{km}$ with two (identical) magnetized iron neutrino detectors (MIND) with a fiducial mass of $50 \, \mathrm{kt}$ each. In addition, a $10 \, \mathrm{kt}$ emulsion cloud chamber (ECC) for $\nu_\tau$ detection is placed at the short baseline.  For each baseline, a total of $2.5 \cdot 10^{21}$ useful muon decays plus $2.5 \cdot 10^{21}$ useful antimuon decays in the straight of the corresponding storage ring is used, which could be achieved by ten years of operation with $2.5 \cdot 10^{20}$ useful muon decays per baseline, year, and polarity. The muon energy $E_\mu$ is assumed to be $25 \, \mathrm{GeV}$, which is sufficient for a detector with a low enough detection threshold~\cite{Huber:2006wb}. The detector and systematics specifications can be found in \Refs~\cite{ids,Autiero:2003fu}. Note that there is not yet any near detector specification. We do not simulate the near detector explicitely, because we do not discuss non-standard production or detection effects such as in \Ref~\cite{Kopp:2007ne}.  In addition, we do not require charge identification in the disappearance channel, which means that we have to add the $\nu_\mu$ and $\bar\nu_\mu$ event rates. It has been demonstrated in \Ref~\cite{Huber:2006wb} that the better efficiencies (and better energy  threshold) lead to a better performance in that case.  
In summary, the following oscillation channels are included:
\begin{enumerate}
\item
 $\nu_e \rightarrow \nu_\mu$ at $4 \, 000 \, \mathrm{km}$ ($\nu_\mu$ appearance)
\item
 $\bar\nu_e \rightarrow \bar\nu_\mu$ at $4 \, 000 \, \mathrm{km}$ ($\bar\nu_\mu$ appearance)
\item
 $\nu_e \rightarrow \nu_\mu$ at $7 \, 500 \, \mathrm{km}$ ($\nu_\mu$ appearance)
\item
 $\bar\nu_e \rightarrow \bar\nu_\mu$ at $7 \, 500 \, \mathrm{km}$ ($\bar\nu_\mu$ appearance)
\item
 $\boldsymbol{\nu_\mu}+\bar\nu_e \rightarrow \boldsymbol{\nu_\mu}+\bar\nu_\mu$ at $4 \, 000 \, \mathrm{km}$ ($\nu_\mu$ disappearance)
\item
 $\boldsymbol{\bar\nu_\mu}+\nu_e \rightarrow \boldsymbol{\bar\nu_\mu}+\nu_\mu$ at $4 \, 000 \, \mathrm{km}$ ($\bar\nu_\mu$ disappearance)
\item
 $\boldsymbol{\nu_\mu}+\bar\nu_e \rightarrow \boldsymbol{\nu_\mu}+\bar\nu_\mu$ at $7 \, 500 \, \mathrm{km}$ ($\nu_\mu$ disappearance)
\item
 $\boldsymbol{\bar\nu_\mu}+\nu_e \rightarrow \boldsymbol{\bar\nu_\mu}+\nu_\mu$ at $7 \, 500 \, \mathrm{km}$ ($\bar\nu_\mu$ disappearance)
\item
 $\nu_e \rightarrow \nu_\tau$  at $4 \, 000 \, \mathrm{km}$ ($\nu_\tau$ appearance)
\end{enumerate}
In the following, we will refer to golden channels~1 to~4 as {\sf Golden}~\cite{Cervera:2000kp},
to channels~5 to~8 as the disappearance channels, and to the silver channel~9 as {\sf Silver}~\cite{Donini:2002rm}.
Note that in the limit of small $\stheta$, channels~5 to~8 can be approximated by $\nu_\mu \rightarrow \nu_\mu$ and $\bar\nu_\mu \rightarrow \bar\nu_\mu$,  respectively, which we have discussed in the phenomenology section.

Compared to \Ref~\cite{ids}, we study several modifications in order to discuss the neutrino factory optimization for non-standard interactions. In some cases, we will vary the muon energy or baseline(s), sometimes even for {\sf Silver} separately. In addition, we will discuss a potentially improved silver channel detector {\sf Silver*}, which uses five times the signal and three times the background of {\sf Silver} in order to implement the hadronic decay channels of the $\tau$ as well~\cite{Huber:2006wb}. In a part of the study, we will not include {\sf Silver} at all.

All simulations are performed using the {\sf GLoBES} software~\cite{Huber:2004ka,Huber:2007ji}.
The experiment description is based on \Refs~\cite{Huber:2002mx,Huber:2006wb} updated with the numbers from \Ref~\cite{ids}. For the true oscillation parameters, we use $\sin^2 \theta_{12}=0.3$,  $\sin^2 \theta_{23}=0.5$, $\sdm=7.9 \, \cdot 10^{-5} \, \mathrm{eV}^2$, $\ldm=2.6 \, \cdot 10^{-3} \, \mathrm{eV}^2$, and a normal mass hierarchy unless stated otherwise.
For the true $\stheta$ and $\deltacp$, we choose certain benchmark points, but we will see that there is relatively little dependence on their true values in most cases. For $\theta_{12}$ and $\sdm$, we assume external measurement precisions of $10\%$ each, whereas we do not impose any external constraints on the leading atmospheric parameters. The used true values and their errors are motivated by the current best-fit values and their errors, see, \eg, \Refs~\cite{Schwetz:2006dh,GonzalezGarcia:2007ib}. For the matter density, we use the PREM profile (Preliminary Reference Earth Model) with a normalization uncertainty of 5\%~\cite{Geller:2001ix,Ohlsson:2003ip}. For neutrino trajectories which do not cross the core of the Earth, we approximate the PREM profile by a single layer of constant density, while for core-crossing neutrinos, we use a mantle-core-mantle profile with three layers. The densities within the respective layers are computed by averaging the full PREM profile along the neutrino trajectory. Note that the 5\% matter density uncertainty is assumed to be correlated among different channels operated at the same baseline, and uncorrelated between different baselines (unless we vary the baseline of one channel independently; in that case, it is always uncorrelated). 

\section{Impact of different channels}
\label{sec:channels}

\begin{figure}[t!]
\begin{center}
\includegraphics[width=\textwidth]{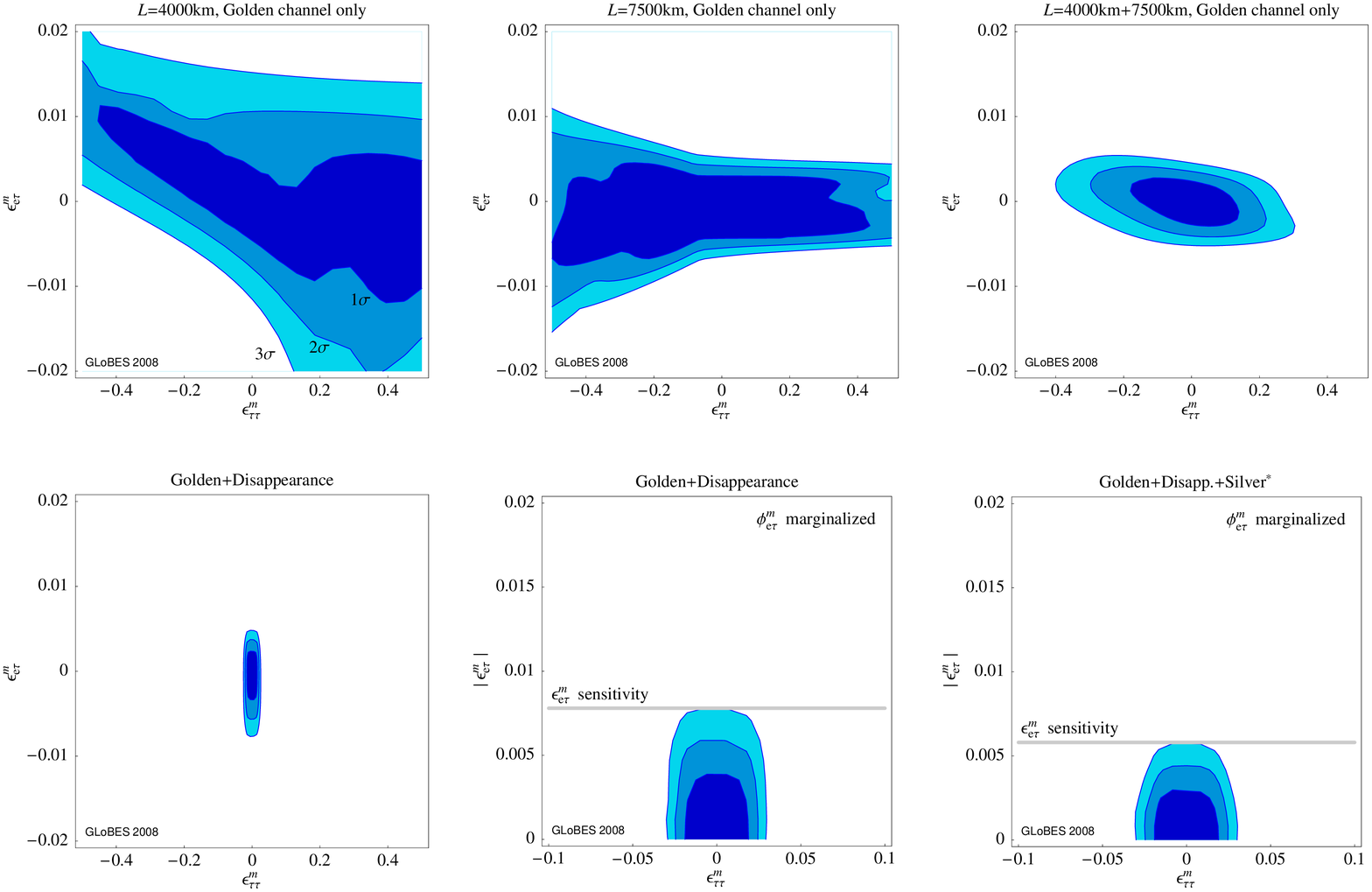}
\end{center}
\mycaption{
\label{fig:channels} Contribution of different channels to the $\ett$-$\eet$ sensitivity.  The first row corresponds to the golden channel only. The different columns correspond to the two different baselines $4 \, 000 \, \mathrm{km}$ and $7 \, 500 \, \mathrm{km}$, as well as their combination. In the second row, we in addition add the disappearance channel (left), introduce complex $\eet$ (middle), and finally add the {\sf Silver*} channel (right). In the upper row, we only marginalize over $\stheta$ and $\deltacp$, whereas in the lower row, we marginalize over all oscillation parameters. 
Note that $\eet$ is assumed to be real in the first four panels, and complex in the last two.
In this figure, a true $\deltacp=3 \pi/2$ and $\stheta=0.001$ have been assumed. In addition, $E_\mu = 50 \, \mathrm{GeV}$ has been chosen for comparison to \Ref~\cite{Ribeiro:2007ud}. The contours correspond to the $1\sigma$, $2\sigma$, and $3 \sigma$ confidence level for 2 d.o.f. 
}
\end{figure}

In this section, we discuss the impact of different oscillation channels, and we study the optimization of the silver channel. We know from \Ref~\cite{Ribeiro:2007ud} that the combination of two baselines, one with about $3 \, 000 \, \mathrm{km}$ and the other with about $7 \, 000 \, \mathrm{km}$, turns out to be very useful to resolve correlations between the standard and non-standard parameters, and among different non-standard parameters. However, disappearance information was not taken into account in the analysis of \Ref~\cite{Ribeiro:2007ud}, and the off-diagonal $\epsilon$'s were assumed to be real.
On the other hand, it has been demonstrated in \Ref~\cite{Kitazawa:2006iq} that the silver channel probability at $3 \, 000 \, \mathrm{km}$ significantly depends on the non-standard effects, especially $\eet$. Therefore, we focus on three major questions in this section:
\begin{enumerate}
\item
 Which oscillation channels dominate the measurements for which non-standard quantities?
\item
 If one has already a two-baseline setup, such as the {\sf IDS-NF} setup, does one still need the
silver channel?
\item
 Is the silver channel location at the shorter of the two golden baselines really the optimal choice?
\end{enumerate}
These questions can only be quantitatively and reliably answered using a full simulation.

To compare our results to \Ref~\cite{Ribeiro:2007ud}, let us first of all assume
all $\epsilon_{\alpha \beta}^m$ to be real.
In addition, we study simultaneous constraints for two non-standard parameters to illustrate the impact of different channels. In order to compare to \Ref~\cite{Ribeiro:2007ud}, we choose an example in the $\ett$-$\eet$ plane, where correlations are particularly severe, \ie , $\sthetat=0.001$ and $\deltacpt=3 \pi/2$. In \figu{channels}
we show the allowed sensitivity region in the upper row for $L=4 \, 000 \, \mathrm{km}$ (left panel), $L=7 \, 500 \, \mathrm{km}$ (middle panel), and the combination of the two baselines (right panel).  In these panels, we have only marginalized over $\stheta$ and $\deltacp$, the true $\eet=\ett$ have been assumed to vanish, and we have chosen $E_\mu=50 \, \mathrm{GeV}$ for the whole figure.\footnote{We will discuss the impact of the muon energy in the next section.} Our results reproduce \Ref~\cite{Ribeiro:2007ud} very well, even though the baselines are slightly changed to match the {\sf IDS-NF} baseline setup. Note that the magic baseline fit is not completely correlation-free, as it is obvious from \equ{nsimb}, which means that there is no clean measurement of $\eet$ at the magic baseline.
In addition, there are still some problems with correlations in the combination of the two baselines. Therefore, one may suspect that the silver channel could help to resolve these. 

We have tested that this problem becomes even worse if one marginalizes over the leading solar and atmospheric oscillation parameters as well. In this case, the silver channel indeed helps to resolve the correlations. However, this conclusion does not hold anymore if one in addition adds the disappearance channel, as we have done in the lower left panel of \figu{channels}. It helps to measure the atmospheric oscillation parameters, and it severely constrains $\ett$ (or, as we have tested, $\emt$). Furthermore, there is almost no correlation remaining between $\ett$ and $\eet$. From the comparison with the upper right panel we learn that  the golden channel indeed has the best $\eet$ sensitivity, whereas the disappearance channel has the best $\emt$ and $\ett$ sensitivity. Therefore, the analytical formulas presented in \Sec~\ref{sec:pheno} are really the ones applicable to the most sensitive channels at the neutrino factory.

Now one can argue that once even more parameters are added, the information from the silver channel needs to contribute at some point. In addition, we have not included the complex phase $\eetp$ yet. Therefore, we show in \figu{channels}, lower middle panel, the full complex case with the additional parameter $\eetp$ marginalized over. Note that the scale on the axes has changed, and that $\ett$ is real by definition. There is no large quantitative change compared to the lower left panel. However, if the {\sf Silver*} channel is in addition used at the $4 \, 000 \, \mathrm{km}$ baseline, as we illustrate in the lower right panel, $\eeta$ can be somewhat better constrained, whereas there is almost no effect for $\ett$ (or $\emt$). Though this improvement is not insignificant (a 30\% effect), we will study in  \Sec~\ref{sec:muonen} how it quantitatively depends on the muon energy and silver channel implementation. We have checked that it cannot be achieved by a mere up-scaling of the {\sf Golden} detectors, as one may naively expect, even if the detector masses are increased by 50~kt each. Therefore, we have identified a synergy in the sense of \Ref~\cite{Huber:2002rs} here. 
We have checked that even if one includes in addition $\emt$ to be marginalized over, there are no significant qualitative changes to this picture, \ie, no additional correlations to be resolved. 
Therefore, we conclude that the {\sf Silver*} channel at the $4 \, 000 \, \mathrm{km}$ baseline is not a key component to push the non-standard parameter measurements by an order of magnitude, but it may help to improve the $\eeta$ sensitivity somewhat. 
One reason are the relatively low event rates even for $\deltacp=3 \pi/2$, where the silver rate becomes largest in the CP-odd term (second term in \equ{so}): in total about $47$ events for $\nu_\tau$ appearance ({\sf Silver*}) or $9$ events for $\nu_\tau$ appearance ({\sf Silver}), compared to about $323$ events for $\nu_\mu$ appearance and 
$6$ million events for $\nu_\mu$ disappearance at the $4 \, 000 \, \mathrm{km}$ baseline ($E_\mu=50 \, \mathrm{GeV}$, $\stheta=0.001$, $\deltacp=3 \pi/2$, normal hierarchy, no NSI). On the other hand, we know from the upper left and middle panels in \figu{channels} that the magic baseline significantly contributes to the $\eet$ measurement for the golden channel in an orthogonal way, and there is still a substantial number of events at this baseline ($126$ events for the above benchmark point). Therefore, we have demonstrated that the  silver channel at the short baseline is not mandatory for the $\eet$ sensitivity if the golden channel at the magic baseline is used. In addition, at the magic baseline, the golden and silver appearance channels are equivalent, as we have found in \equ{nsimb}.
Hence, we do not expect the silver channel to be useful at the magic baseline either. 

\begin{figure}[t]
\includegraphics[width=\textwidth]{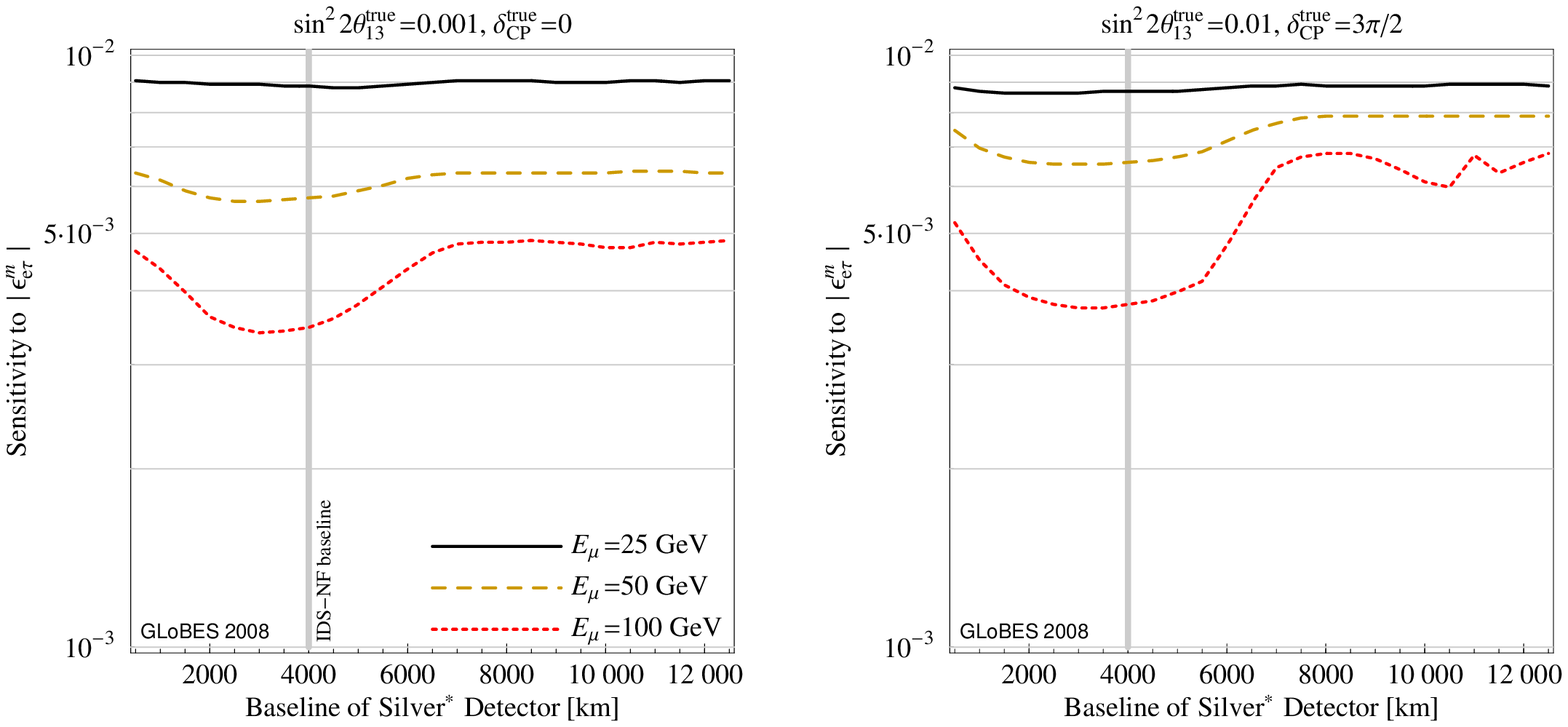}
\mycaption{\label{fig:silveropt}
Sensitivity to $\eeta$ ($3 \sigma$) as a function of the {\sf Silver*} baseline for three detector setups: 
two MIND detectors at $4 \, 000 \, \mathrm{km}$ and $7 \, 500 \, \mathrm{km}$,
and one {\sf Silver*} detector at the specified baseline on the horizontal axes. The different curves correspond to different muon energies as given in the plots. The different panels show the result for different (representative) true values of $\stheta$ and $\deltacp$, as given in the captions.  The {\sf IDS-NF} iron detector short baseline is marked by the vertical lines.}
\end{figure}

Even if the silver channel does not help at $4 \, 000 \, \mathrm{km}$ or $7 \, 500 \, \mathrm{km}$, what about (hypothetically) placing the ECC at a third baseline in combination with two {\sf Golden} detectors at $4 \, 000 \, \mathrm{km}$ and $7 \, 500 \, \mathrm{km}$? We discuss this question for $\eeta$ in \figu{silveropt} for two (representative) sets of $\sthetat$ and $\deltacpt$. The different curves correspond to the different muon energies $25 \, \mathrm{GeV}$, $50 \, \mathrm{GeV}$, and $100 \, \mathrm{GeV}$. Note that in this case, the matter density is assumed to be uncorrelated among all three baselines. Indeed one can read off these figures that the optimal performance is obtained at 
about $3 \, 000 \, \mathrm{km}$ to $4 \, 000 \, \mathrm{km}$, where the vertical lines correspond to our standard choice. This means that our setup is perfectly optimized for the silver channel.
In addition, one can read off this figure that for $E_\mu \ll 50 \, \mathrm{GeV}$ hardly any effect is visible.
We will discuss the muon energy dependence in detail in the next section.

Note that we will not consider two NSI parameter correlations for the rest of this study anymore, since we have found that there is hardly any correlation remaining in the $\ett$-$\eet$ plane (\cf, \figu{channels}, lower middle and right panels). This means that the sensitivities can as well be studied separately.\footnote{In fact, for very large $\stheta$ close to the current bound, there is some correlation in the $\ett$-$\eet$ plane remaining, which partly comes from the matter density uncertainty. Since for $\stheta \gtrsim 0.01$, a neutrino factory would probably look different from the current {\sf IDS-NF} baseline setup (\ie, have a short baseline and a lower muon energy)~\cite{Geer:2007kn,Huber:2007uj,Bross:2007ts}, we do not discuss this case anymore.} Of course, there is a straightforward correlation in the $\emt$-$\ett$ plane, which can be analytically understood from the full two-parameter mapping~\cite{Blennow:2005qj}.  For example, for maximal atmospheric mixing and real $\emt$, we have (\cf, \Sec~\ref{sec:ettemt} and \Ref~\cite{Blennow:2005qj})
\begin{equation}
(2 \hat{A} \emt + 1)^2 + (\hat{A} \ett)^2 = \left( \frac{\Delta \tilde{m}_{31}^2}{\ldm} \right)^2 =  const. \, ,
\end{equation}
which just corresponds to the normal form of an ellipse centered at $\emt = -1/(2 \hat{A})$ and $\ett=0$.
We therefore do not consider this correlation anymore. The $\emt$-$\eet$ plane, on the other hand, is similar to the $\ett$-$\eet$ plane, as it is obvious from \Sec~\ref{sec:ettemt}. In principle, correlations between $\eee$ and the other parameters could be interesting, but this would be out of the main line of this study. Note that the correlation with $\eee$ is intimately connected to the matter density uncertainty, which we have taken into account. This means that our simulation using a 5\% matter density uncertainty is, apart from the fact that the matter density is assumed to be uncorrelated between the two baselines, equivalent to a simulation with a precisely known matter density profile and an external bound $\eeea \lesssim 0.05$ ($1\sigma$). 
We will further comment on the relationship to the matter density uncertainty in \Sec~\ref{sec:muonen}.

\section{Optimal muon energy}
\label{sec:muonen}

\begin{figure}[t!]
\begin{center}
\includegraphics[width=\textwidth]{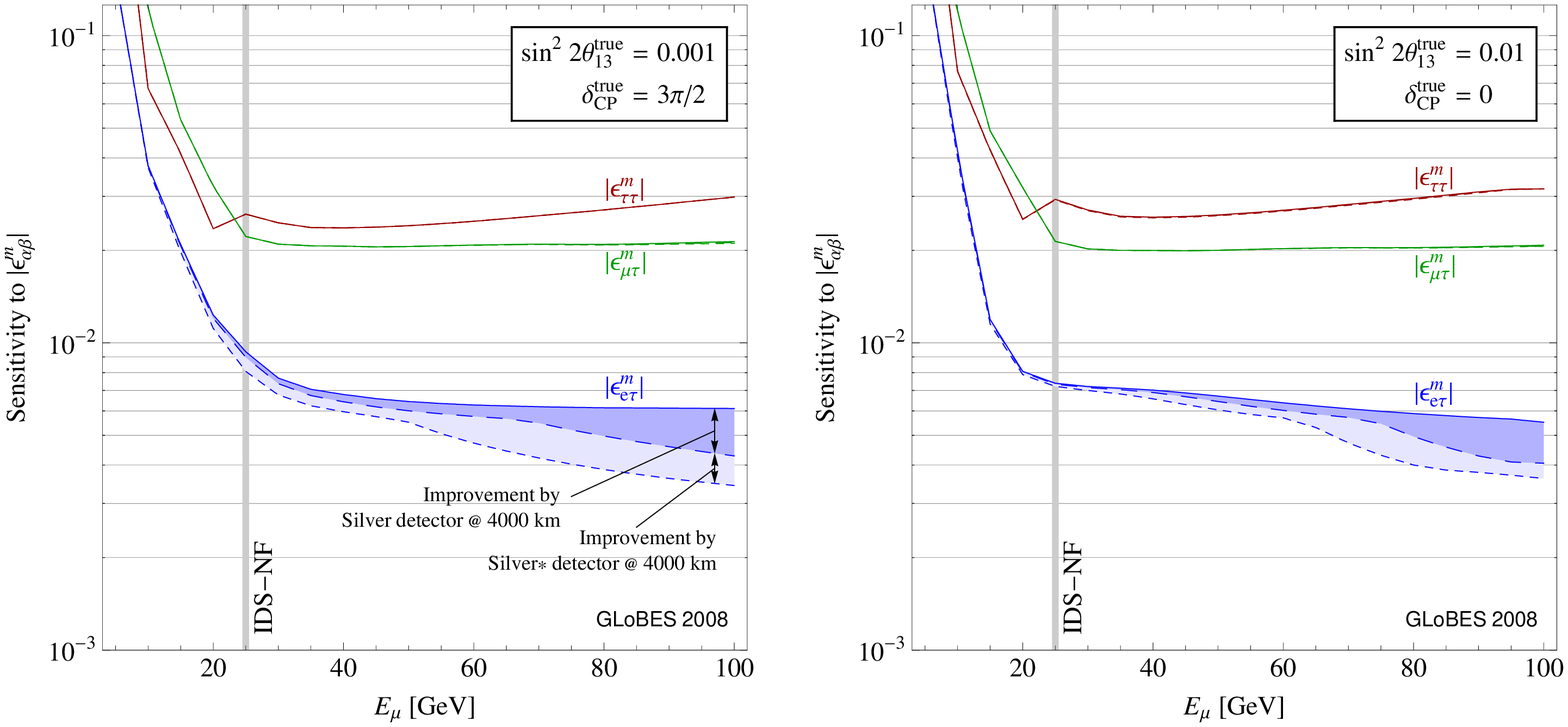}
\end{center}
\mycaption{\label{fig:emudep}
Sensitivity to $\eeta$, $\emta$, and $\etta$  as a function of the muon energy for different detector setups ($3 \sigma$): 
two magnetized iron detectors at $4 \, 000 \, \mathrm{km}$ and $7 \, 500 \, \mathrm{km}$,
and one optional {\sf Silver} or {\sf Silver*} detector at the shorter baseline. The two panels show the result for different (representative) true values of $\stheta$ and $\deltacp$, as given in the plots. Here a normal mass hierarchy is assumed. The {\sf IDS-NF} standard muon energy ($25 \, \mathrm{GeV}$) is marked by vertical lines.}
\end{figure}

From \figu{silveropt}, we have learned that the muon energy has some impact on the NSI performance. 
This can, for example, be seen at the relative dominance of the NSI terms for higher neutrino energies in \equ{nsimb}. Therefore, we discuss in this section the dependence of the NSI parameter sensitivities on the muon energy, with and without silver channel. Note that for the standard oscillation parameters, the muon energy has only a minor impact if the detection threshold in the {\sf Golden} detectors is low enough and $E_\mu \gtrsim 20 \, \mathrm{GeV}$~\cite{Huber:2006wb}. Therefore, a lower muon energy $E_\mu=25 \, \mathrm{GeV}$ was chosen for the {\sf IDS-NF} baseline setup than the originally anticipated $E_\mu=50 \, \mathrm{GeV}$.
For a possibly different detector technology, also a low energy
neutrino factory with $E_\mu \sim 5 \, \mathrm{GeV}$ has been
discussed in the literature~\cite{Geer:2007kn,Huber:2007uj,Bross:2007ts}.
Such an experiment may be useful for large $\stheta$, but the typical
setups involve only one baseline, and will therefore not be considered
in this work.
Again, there are three relevant questions for this section:
\begin{enumerate}
\item
 Is the muon energy of $25 \, \mathrm{GeV}$ sufficient for the NSI sensitivities, or should one go to a higher $E_\mu$ for the {\sf IDS-NF} baseline?
\item
 What are the prospects to improve the current NSI bounds for a considerably lower muon energy?
\item
 How important is the silver channel as a function of the muon energy?
\end{enumerate}
In order to address these questions, we present in \figu{emudep} the sensitivity to $\eeta$, $\emta$, and $\etta$ ($3 \sigma$) as a function of the muon energy for two different sets of true values. The different curves correspond to different $|\epsilon^m_{\alpha \beta}|$ and different detector setups: 
two magnetized iron detectors at $4 \, 000 \, \mathrm{km}$ and $7 \, 500 \, \mathrm{km}$,
and one optional {\sf Silver} or {\sf Silver*} detector at the shorter baseline. For $\eeta$, we find some dependence on $\sthetat$ and $\deltacpt$, which, however, does not lead to qualitatively different conclusions.  For $\emt$ and $\ett$, however, we hardly find any dependence on $\sthetat$ and $\deltacpt$.

Neglecting the silver channel for the moment, we find in all cases in \figu{emudep} a strong depletion of the sensitivity for $E_\mu \lesssim 20 \, \mathrm{GeV}$. For higher muon energies, however, there is no significant gain anymore.  The reason is that the energy range with the strongest matter effects is sufficiently covered, while for higher muon energies, the event rates at the lower end of the spectrum decrease somewhat. In fact, the $\etta$ sensitivity even becomes worse for $E_\mu \gtrsim 50 \, \mathrm{GeV}$. This means that $E_\mu = 25 \, \mathrm{GeV}$ is indeed sufficient for the NSI sensitivities. For $E_\mu \simeq 5 \, \mathrm{GeV}$, however, the current bounds could only be improved for $\eeta$ and $\etta$ by a factor of a few, and not at all for $\emta$. Note that also the standard oscillation parameter measurements are significantly affected for $E_\mu \ll 20 \, \mathrm{GeV}$~\cite{Huber:2006wb}.

For the silver channel, we do not find any impact for the $\emta$ and $\etta$ sensitivities whatever the chosen muon energy is.
For the $\eeta$ sensitivity, however, it slightly improves the sensitivity for $E_\mu \gtrsim 25 \, \mathrm{GeV}$, depending on the true parameter values. For $E_\mu=25 \, \mathrm{GeV}$, the silver channel hardly contributes. This means that, at least for the NSI, the current choice of $E_\mu = 25 \, \mathrm{GeV}$ for the  {\sf IDS-NF} baseline setup is in contradiction with the {\sf Silver} detector at $4 \, 000 \, \mathrm{km}$. Either the muon energy needs to be increased to make the silver channel valuable, or there is no physics case for the silver detector. Since the impact of the silver channel is, even for $E_\mu=50 \, \mathrm{GeV}$, not very large, we prefer to choose $E_\mu=25 \, \mathrm{GeV}$ in the following in order to be consistent with the {\sf IDS-NF} baseline setup. However, we will not include the {\sf Silver} detector, because we have not found any significant physics contribution. 

\begin{figure}[t!]
\begin{center}
\includegraphics[width=\textwidth]{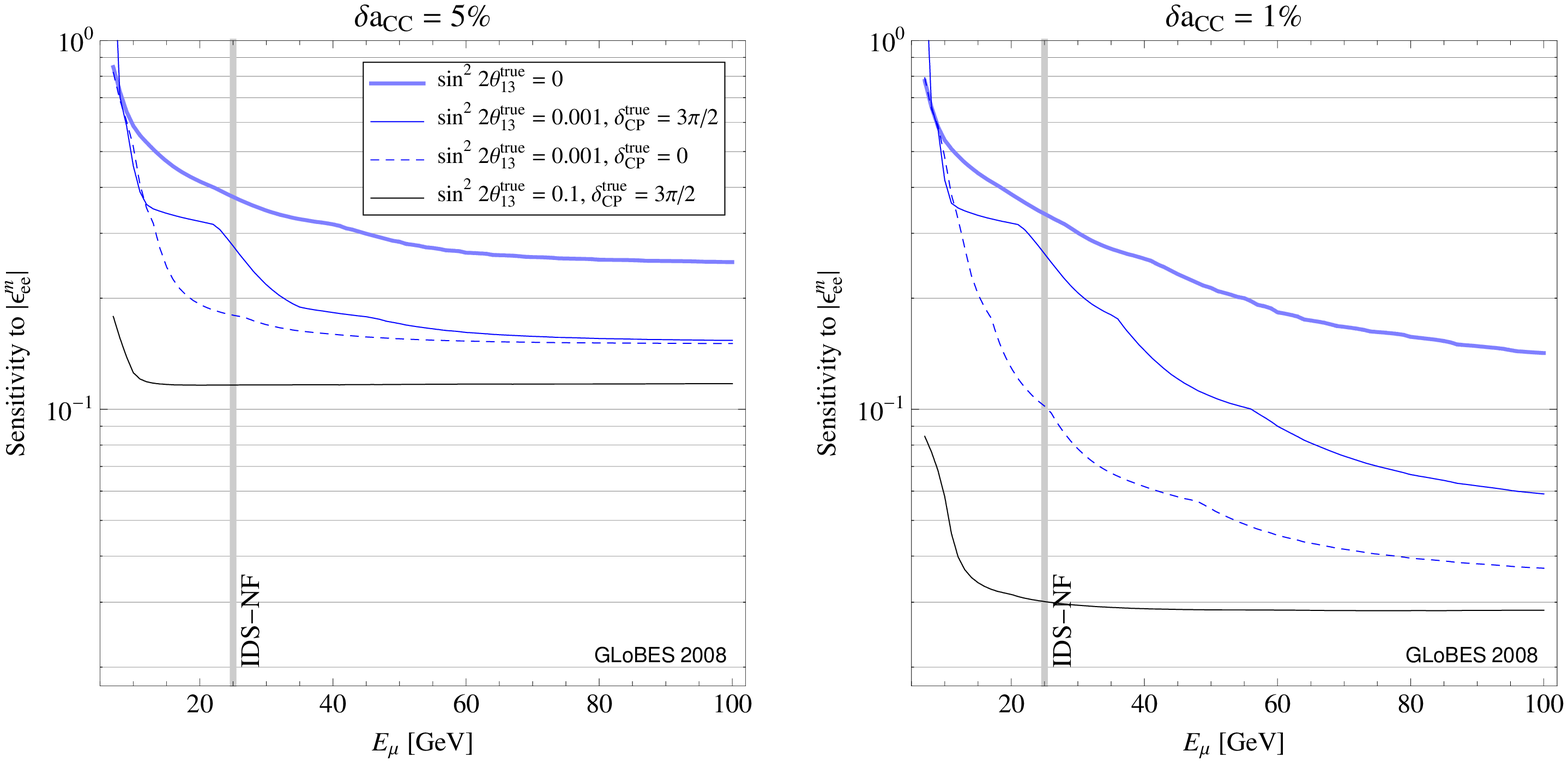}
\end{center}
\mycaption{\label{fig:emudepee}
Sensitivity to $\eeea$  as a function of the muon energy for two magnetized iron detectors at $4 \, 000 \, \mathrm{km}$ and $7 \, 500 \, \mathrm{km}$ ($3 \sigma$). The left panel is computed for a matter density uncertainty $\delta a_{\mathrm{CC}}= 5\%$, and the right panel for a matter density uncertainty  $\delta a_{\mathrm{CC}}=1\%$ ($1\sigma$). Here different values for $\sthetat$ and  $\deltacpt$ have been used, as shown in the plot legend. The {\sf IDS-NF} standard muon energy ($25 \, \mathrm{GeV}$) is marked by vertical lines.}
\end{figure}

As far as the $\eeea$ sensitivity is concerned, let us first of all make some numerical estimates. From \equ{Ham},
we know that $a_{\mathrm{CC}} +a_{\mathrm{CC}} \, \eee$ enters the Hamiltonian in the presence of $\eee$. If, in addition, a matter density shift $x$ is considered, \ie, $a_{\mathrm{CC}} \rightarrow a_{\mathrm{CC}} + x$, the matter potential $a_{\mathrm{CC}}$ is (to leading order) shifted by $P=x + a_{\mathrm{CC}} \, \eee$. For $\eee=0$, \ie, without NSI, $P = x$ is exactly the deviation of the matter density from the reference matter density profile. Note that we impose a prior on the shift $x$, which means that $|x| \lesssim \delta a_{\mathrm{CC}}$ is limited, where $\delta a_{\mathrm{CC}}$ is the matter density uncertainty ($1\sigma$).
This uncertainty comes, for example, from the limited precision of seismic wave experiment.
The measurement of $P$, \ie, the matter density precision measurement, has been studied in the literature in \Refs~\cite{Winter:2005we,Minakata:2006am,Gandhi:2006gu}.
Therefore, one can, in principle, estimate the $\eee$ sensitivity from the precision of $P$ and $\delta a_{\mathrm{CC}}$ as the Gaussian average for a single baseline experiment. For two baselines, $\eee$ will be correlated between the baselines, whereas $x$ will be not. Therefore, a slightly better $\eee$ sensitivity might be expected in practice.

We show in \figu{emudepee} the sensitivity to $\eeea$  as a function of the muon energy for two magnetized iron detectors at $4 \, 000 \, \mathrm{km}$ and $7 \, 500 \, \mathrm{km}$ ($3 \sigma$). The left panel is computed for a matter density uncertainty $\delta a_{\mathrm{CC}}=5\%$, and the right panel for a matter density uncertainty $\delta a_{\mathrm{CC}}= 1\%$ ($1\sigma$). Here different values for $\sthetat$ and  $\deltacpt$ have been used, as shown in the plot legend. 
For very high $E_\mu$, the effective precision on $P$ will be higher than the matter density uncertainty $\delta a_{\mathrm{CC}}$, which means that the $\eeea$ sensitivity will be asymptotically limited by $\delta a_{\mathrm{CC}}$. 
Therefore, in the left panel of \figu{emudepee}, the sensitivities are roughly limited by $3 \cdot 0.05/\sqrt{2} \simeq 0.11$, in the right panel by $3 \cdot 0.01/\sqrt{2} \simeq 0.02$, where the factor $\sqrt{2}$ comes from the fact that two independent matter density priors are added. For very low $E_\mu$, the precision of $P$ will be much weaker than $\delta a_{\mathrm{CC}}$ in all cases, which means that the sensitivity is limited by the precision of $P$. Therefore, the curves in both panels are very similar for small muon energies irrespective of the matter density uncertainty. 
The dependence on $\sthetat$ is similar to that of the measurement of $P$ without NSI, see \Ref~\cite{Minakata:2006am}. Note that there is hardly any dependence on $\deltacpt$ for very small or very large $\sthetat$. 
For very large $\stheta$, the precision of $P$ is extremely good already for comparatively small $E_\mu$, which means that the asymptotic limit is quickly reached (\cf, curves for $\stheta=0.1$).
For very small $\sthetat$, the solar term (fourth term in \equ{so}) dominates the measurement, which means that the performance is poorer than in the large $\stheta$ limit.\footnote{In fact, the fourth term in \equ{so} is CP-invariant, which means that it is also invariant under $\eee = 0 \rightarrow \eee = -2$ (which corresponds to a sign flip of the matter density profile). We have not included this additional degeneracy.}
For intermediate $\sthetat$, the performance strongly depends on $\deltacpt$, because $\deltacp$ leads to non-trivial correlations (\cf, curves for $\sthetat=0.001$).
Compared to the other sensitivities discussed in this section, there can be a strong gradient between a muon energy of 25 and~50~GeV especially if $\delta a_{\mathrm{CC}}$ is sufficiently small, which, however, somewhat depends on the true parameters. Therefore, if one emphasizes the $\eeea$ sensitivity, a higher muon energy might be important.  Since the improvement would only be a factor of a few beyond the current bounds, this would probably not be the main argument for a higher muon energy.

\section{Baseline optimization: Standard versus non-standard physics}
\label{sec:baselines}

In this section, we discuss the optimization of the baselines for a
neutrino factory with two detectors, considering both standard oscillation
physics and non-standard scenarios. The experimental setup is based
on the {\sf IDS-NF}~1.0 configuration, but omitting the silver channel.
We will treat the two baselines $L_1$ and $L_2$ as free parameters
in this section, and compare their resulting optimal values to
the ones suggested by the {\sf IDS-NF}~1.0 setup, namely $L_1 \sim 3 \, 000 \, \mathrm{km}$
to $5 \, 000 \, \mathrm{km}$, and $L_2 \sim 7 \, 000 \, \mathrm{km}$
to $8 \, 000 \, \mathrm{km}$. For definiteness, we define our benchmark
setup by $L_1 = 4 \, 000 \, \mathrm{km}$ and $L_2 = 7 \, 500 \, \mathrm{km}$.
This choice is based on previous works, in particular on the magic baseline
argument in \Ref~\cite{Huber:2003ak} (correlations and degeneracies disappear
at a baseline around $7 \, 500 \, \mathrm{km}$), the single baseline
optimization in \Ref~\cite{Huber:2006wb} (leading to the conclusion
that one needs two baselines to optimally measure all standard performance
indicators), and the optimization of the longer baseline $L_2$ with the
shorter one fixed at $L_1=4 \, 000 \, \mathrm{km}$ in \Ref~\cite{Gandhi:2006gu}.
A simultaneous variation of both baselines has so far only been considered
in \Ref~\cite{Huber:2003ak} for the $\stheta$ sensitivity reach.

The main questions relevant for this section therefore are:
\begin{enumerate}
\item
 Is the {\sf IDS-NF} baseline setup still optimal for {\em all} standard
oscillation performance indicators if both baselines are varied simultaneously?
\item   
 Is the standard optimization robust if there are NSI?
\item
 What would the optimal baselines be for non-standard interactions?
\item
 Is the non-standard optimization consistent with the standard one?
\end{enumerate}

\subsection*{Optimization for standard oscillation performance indicators}

\begin{figure}[tp]
  \begin{center}
    \includegraphics[width=\textwidth]{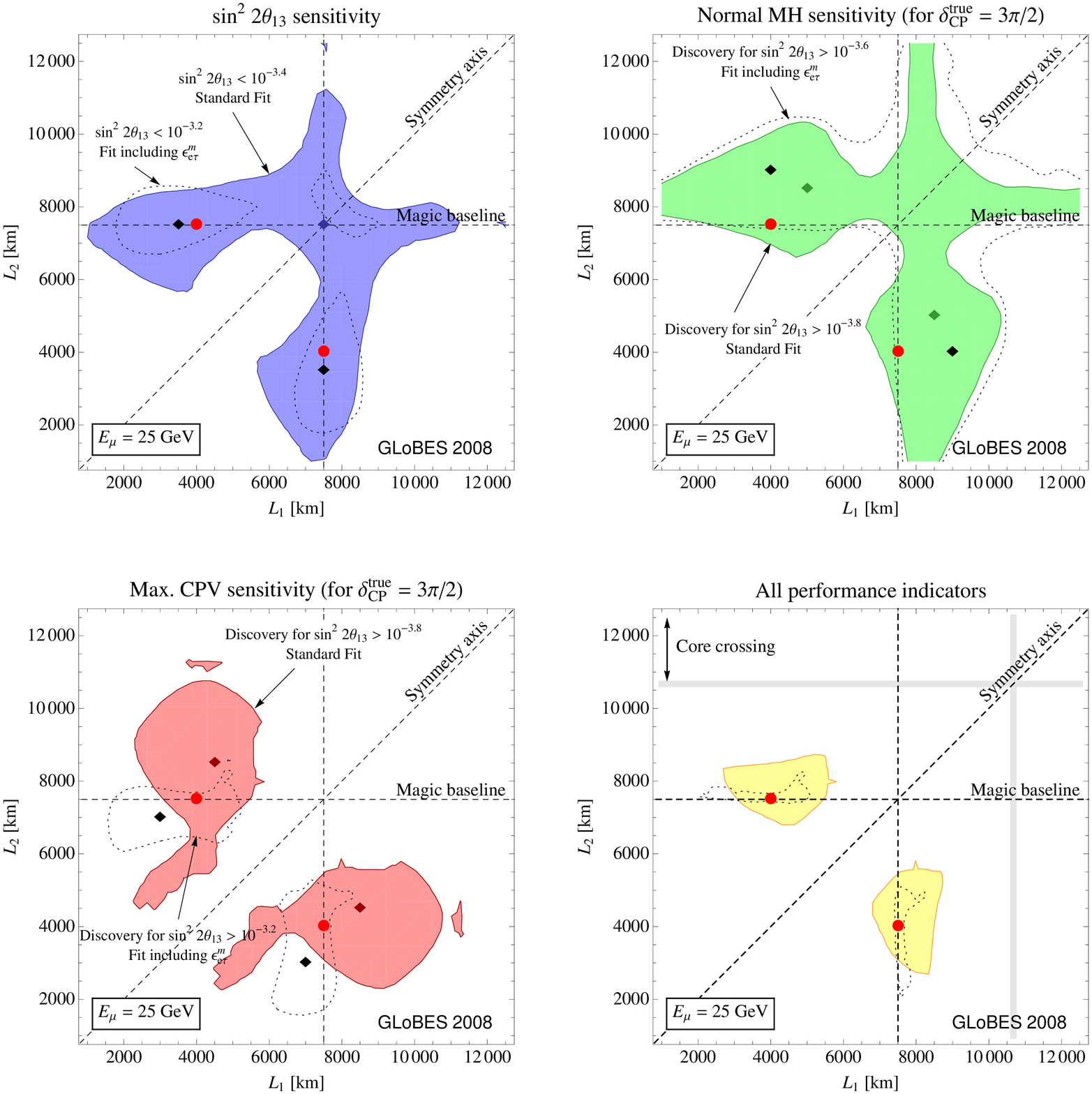}
  \end{center}
  \mycaption{  \label{fig:allopt}
Two-baseline optimization of a neutrino factory (with two {\sf Golden}
    detectors) for the standard oscillation performance indicators. The upper left panel shows the
    (shaded) region where the sensitivity to $\stheta$ better than $10^{-3.4}$ ($5 \sigma$),
    the upper right panel shows the (shaded) region where the sensitivity to the normal mass hierarchy (MH)
    is given for all $\stheta \ge 10^{-3.8}$ ($5 \sigma$, $\delta_{\rm CP}^{\rm true} = 3\pi/2$),  
 the lower left panel shows the (shaded) region where the sensitivity to maximal CP violation (CPV)
    is given for all $\stheta \ge 10^{-3.8}$ ($5 \sigma$, $\delta_{\rm CP}^{\rm true} = 3\pi/2$),  
   and the lower right panel shows the intersection of the three regions as the shaded region.
 The dotted curves have been obtained from a fit including
    $\eet$ marginalized (for the $\stheta$ ranges given in the plots).
    The diamonds show the setups with optimal sensitivities (colored/gray for the shaded contours, black for
the dotted contours),  whereas the circles correspond to the {\sf IDS-NF}
    standard choices $L_1 = 4 \, 000$~km and $L_2 = 7 \, 500$~km. }
\end{figure}

In order to discuss the optimization for the standard performance indicators,
we show in \figu{allopt} the two-baseline optimization of a neutrino factory with two {\sf Golden}
    detectors. In this figure, optimal performance means optimal reach in $\stheta$.
The upper left panel shows the region with optimal sensitivity to $\stheta$, the upper right panel shows the region with optimal sensitivity to the normal mass hierarchy, 
 the lower left panel shows the region with optimal sensitivity to maximal CP violation,
and the lower right panel shows the intersection of the three regions.
The contours have been chosen such that all regions are of similar size
(see values in plots). All performance indicators are defined at the
$5 \sigma$ confidence level in order to include all degeneracies
(even if they occur only at a relatively high $\chi^2$). For the sensitivity
to the mass hierarchy and to CP violation, we use $\delta_{\rm CP}^{\rm true}
= 3\pi/2$, since for this value, degeneracies have a strong impact, so
that it corresponds to a conservative assumption. The circles mark our
{\sf IDS-NF} benchmark setup, whereas the colored (gray) diamonds mark the
optimum baselines for each case. The dotted curves and the black diamonds
indicate the optimization in a non-standard scenario, which will be discussed
below.

Let us, however, first focus on the standard optimization only. For the $\stheta$ sensitivity (upper left panel), we recover the shape from \Ref~\cite{Huber:2003ak}, Fig.~1. In this case, almost any combination of baselines, with one of them being magic, leads to a good performance. The same conclusion can be obtained for the mass hierarchy sensitivity (upper right panel), where the the optimal baseline somewhat varies with $\deltacpt$ (\cf, \Ref~\cite{Huber:2006wb}). Note that for $\stheta$ as well as for the mass hierarchy, the long baseline is a prerequisite. For CP violation (lower left panel), however, one of the baselines has to be short, \ie, between $2 \, 000 \, \mathrm{km}$ and $6 \, 000 \, \mathrm{km}$, while a long baseline is required to resolve the degeneracies (for $\deltacpt=\pi/2$ one might not need that~\cite{Huber:2006wb}). Interestingly, no points on the diagonal are within the optimal region, which means that it is not sufficient to use only one baseline for this set of parameters.
Finally, the lower right panel shows the intersection of the other regions. It clearly demonstrates that our standard choice, denoted by the circle, is well within the optimal region for all performance indicators.

As the next step, assume that the non-standard effects are taken into account in the fit, which will obviously spoil the
standard oscillation parameter sensitivities of a neutrino factory (see, \eg, \Refs~\cite{Huber:2001de,Huber:2002bi}).
This means that we have to marginalize over these effects as well. Let us first of all focus on $\eet$: The dotted curves in \figu{allopt} have been obtained from a fit where also $\eet$ (its absolute value and phase) has been marginalized over. As one can read off the contours, the absolute performances for all the standard oscillation parameters become worse. However, the optimization does not change, as it can be read off from the lower right panel. Therefore, our two-baseline optimized setup is very robust even with respect to non-standard $\eet$.
We have also checked the other $\epsilon_{\alpha \beta}^m$'s for this optimization. While $\emm$, $\emt$, and $\ett$ hardly have any effect on the appearance channel at all (see also analytical formulas in \Ref~\cite{Kopp:2007ne}),
$\eem$ has a similar qualitative effect as $\eet$. 
However, if the stringent existing bounds on $\eema$ are taken into account,
the possible effects of this parameter become completely negligible.

\subsection*{Optimization for the non-standard sensitivities}

\begin{figure}[p!]
  \begin{center}
    \includegraphics[height=0.8\textheight]{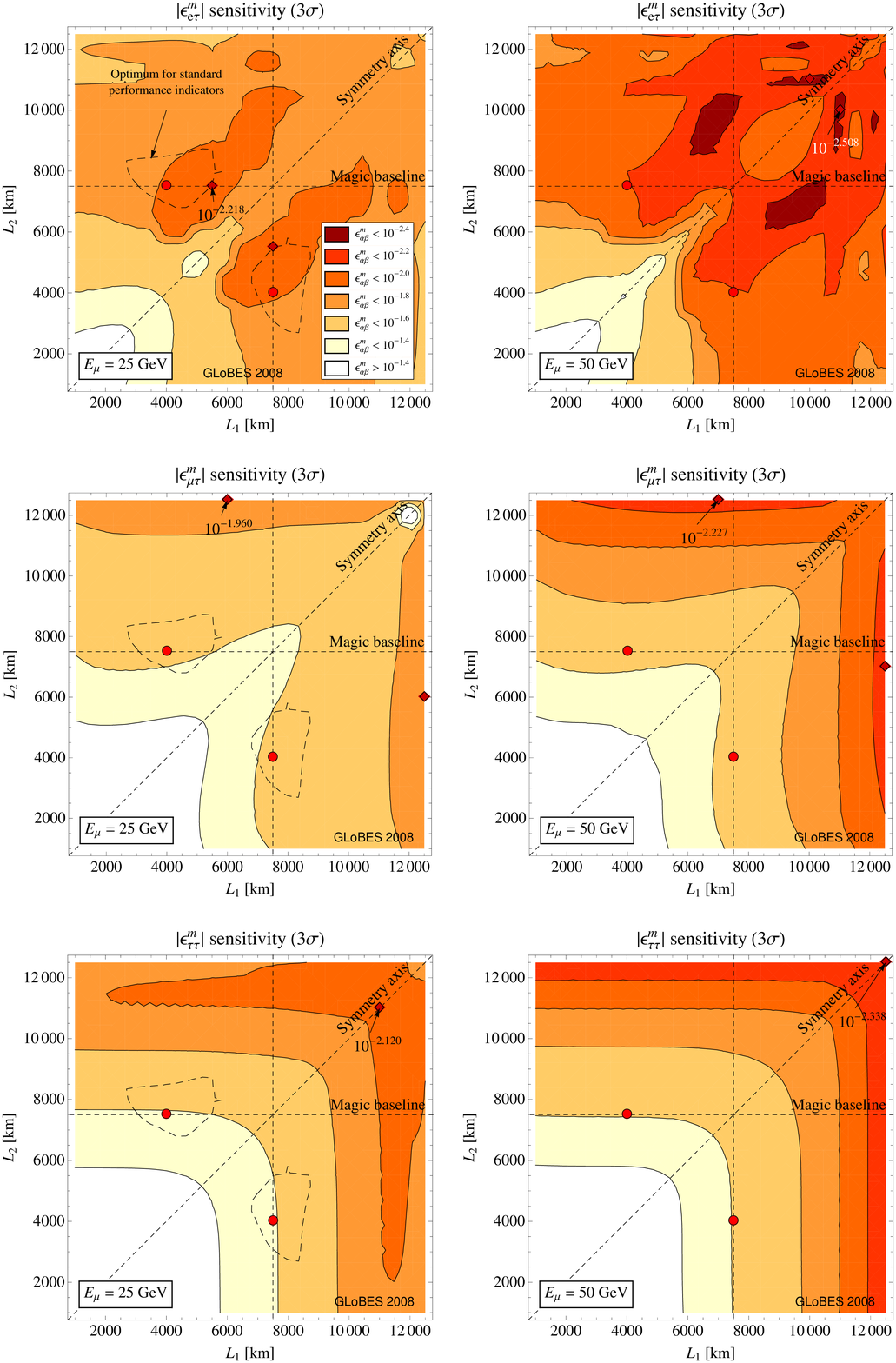}
  \end{center}
  \mycaption{\label{fig:alloptNSI}Two-baseline optimization of a neutrino factory (with two {\sf Golden}
    detectors) for non-standard interactions. The different rows represent the sensitivities to the
   non-standard interaction  parameters $\eeta$, $\emta$,
    and $\etta$, respectively ($3 \sigma$), whereas the different columns represent $E_\mu = 25 \, \mathrm{GeV}$ (left)
    and $E_\mu = 50 \, \mathrm{GeV}$ (right). The dashed curves show the optimum configurations for the
    standard optimization (shaded region from the lower right panel of \figu{allopt}). 
    True parameter values of $\sin^2 2\theta_{13}^{\rm true}
    = 10^{-3}$ and $\delta_{\rm CP}^{\rm true} = 3\pi/2$ have been
    assumed. The diamonds show the
    optimal configurations for non-standard physics, 
whereas the circles correspond to the {\sf IDS-NF} standard choices $L_1 = 4 \, 000$~km and $L_2 = 7 \, 500$~km.}
\end{figure}

We discuss the two-baseline optimization of a neutrino factory (with two {\sf Golden}
    detectors) for non-standard interactions in \figu{alloptNSI}. In this figure, the different rows represent the
   non-standard interaction  parameters $\eeta$, $\emta$,     and $\etta$, respectively, whereas the different columns represent $E_\mu = 25 \, \mathrm{GeV}$ (left)     and $E_\mu = 50 \, \mathrm{GeV}$ (right).
Let us focus on $E_\mu=25 \, \mathrm{GeV}$ (left column) first, which is the {\sf IDS-NF} choice.
In these figures, the dashed curves show the optimum configurations for the
    standard optimization (shaded region from the lower right panel of \figu{allopt})
for comparison. For the $\eeta$ sensitivity (upper left panel), the {\sf IDS-NF} baseline combination $4 \, 000 \, \mathrm{km}$ plus $7 \, 500 \, \mathrm{km}$ is close to optimal for both the standard (circles) and non-standard (diamonds) sensitivities. In this case, slightly longer rather than shorter baselines are preferred. Note that here the main contribution comes from the appearance channels. For the $\emta$ and $\etta$ sensitivities (middle and lower left panels), respectively, much longer baselines are preferred, such as core crossing baselines $L \gtrsim 10 \, 700 \, \mathrm{km}$. The standard and non-standard optimizations do not coincide, which means that one would (hypothetically) need a third baseline. Here the main contribution comes from the disappearance channels, which tend to perform better at long baselines because more oscillation nodes can be resolved. 

In the right column of \figu{alloptNSI}, we increase $E_\mu$ to $50 \, \mathrm{GeV}$. While the absolute sensitivities at the standard choice $4 \, 000 \, \mathrm{km}$ plus $7 \, 500 \, \mathrm{km}$ remain almost unaffected, better absolute sensitivities can be obtained for longer baselines. For example, for the $\eeta$ sensitivity, the bound could be improved by about a factor of two if one went to a different (longer baseline) combination. The reason are the higher event rates for $E_\mu=50 \, \mathrm{GeV}$, which allow for better statistics at even longer baselines. Note, however, that the optimal region for $\eet$ somewhat depends on the chosen $\sthetat$ and $\deltacpt$, whereas the ones for $\emt$ and $\ett$ are almost independent of these parameters. Since the qualitative discussion does not change, we decided to present the results only for one set of parameters.
In addition, we have checked the optimization for the $\eema$ and the $\emma$ sensitivities.  For the $\emma$ sensitivity, there are hardly any qualitative and quantitative changes compared to the $\etta$ sensitivity (see also analytical discussion in \Sec~\ref{sec:ettemt}). For the $\eem$ sensitivity, however, one baseline should be rather short $2 \, 000 \, \mathrm{km} \lesssim L \lesssim 4 \, 000 \, \mathrm{km}$ whereas the other can be long. This means that the standard {\sf IDS-NF} baseline choice is close-to-optimal for this sensitivity. However, the optimal absolute sensitivity is about $0.005$ ($3 \sigma$), \ie, not better than the current bounds, which means that this aspect is of little relevance.

In summary, at least one very long baseline is an important prerequisite to put stronger bounds on the non-standard interactions. For $E_\mu=25 \, \mathrm{GeV}$, the standard and $\eeta$ optimizations are consistent, while $\emta$ and $\etta$ prefer one baseline to be as long as possible. However, even better absolute sensitivities could be achieved for longer baselines in combination with a higher muon energy.

\section{Summary and discussion}
\label{sec:summary}

We have discussed the optimization of a neutrino factory for non-standard interactions (NSI) in the neutrino propagation in terms of  muon energy, baselines, and oscillation channels. Our study has been based on both analytical formulas, and a full simulation of the {\sf IDS-NF} (international design study of a neutrino factory) baseline setup with {\sf GLoBES}. We have considered all possible non-standard parameters $\epsilon_{\alpha \beta}^m$, and have also included the complex phases of the off-diagonal elements.

As far as the different $\epsilon_{\alpha \beta}^m$ and different oscillation channels are concerned, we have identified the $\nu_\mu$ appearance channel as the main contribution to the $\eeta$ sensitivity and the $\nu_\mu$ disappearance channel as the main contribution to the $\emta$ and $\etta$ sensitivities. Furthermore, $\eema$ and $\emma$ cannot significantly be constrained beyond the current bounds at the neutrino factory (\cf, \Tab~\ref{tab:summary}), and the $\eeea$ sensitivity can be directly related to the matter density precision measurement dominated by the $\nu_\mu$ appearance channel. Therefore, we have focused on $\eet$, $\emt$, and $\ett$ in the main line of this study. For example, we have presented analytical formulas for these quantities and the corresponding channels.
Note that we have only considered one non-standard parameter at a time, because we have demonstrated that the two-parameter correlations are either unimportant if appearance and disappearance information is used, such as in the $\ett$-$\eet$ or $\emt$-$\eet$ planes (or $\eee$-$\emt$, $\eee$-$\ett$ planes), they can be analytically understood in a straightforward way, such as in the $\emt$-$\ett$ plane, or they can be related to the matter density uncertainty, such as in the $\eee$-$\eet$ plane.

We have also considered the silver $\nu_e \rightarrow \nu_\tau$ channel for non-standard interactions, and we have only found a synergistic, but small contribution to the $\eeta$ sensitivity if $E_\mu \gtrsim 25 \, \mathrm{GeV}$. This finding is in tension with the current {\sf IDS-NF} baseline setup: At least for NSI, if the muon energy is chosen to be as low as $25 \, \mathrm{GeV}$, the silver channel will be hardly useful.  In combination with the standard oscillation parameter optimization from \Ref~\cite{Huber:2006wb}, we conclude that the tension can only be released if either the muon energy is increased, or if the emulsion cloud chamber is removed from the {\sf IDS-NF} setup.
Except for the silver channel contribution, we have demonstrated that the NSI sensitivities do not significantly improve anymore as a function of the muon energy if $E_\mu \gtrsim 25 \, \mathrm{GeV}$, unless $\eee$ is searched for at intermediate $\sthetat \simeq 0.001$ in well-known matter density environments.

Furthermore, we have revisited the optimization for the standard oscillation parameters as a function of the two baselines of the two main detectors, and we have found that the optimal detector locations are consistent with the {\sf IDS-NF} setup. We have then established the robustness of this optimization with respect to a possible NSI pollution, even though the absolute sensitivities become deteriorated.
As the next step, we have studied the NSI sensitivities as a function of the two baselines. We have found that the optimization of the standard oscillation parameters is consistent with the one for $\eeta$, while for $\emta$ and $\etta$, even longer baselines are in principle preferred (basically, as long as possible within the Earth's diameter). Note that in all cases, one very long baseline ($\gtrsim 7\,000$~km) has turned out to be a key component for the non-standard matter effect measurements. 

\begin{figure}[tp!]
  \begin{center}
    \includegraphics[width=0.9\textwidth]{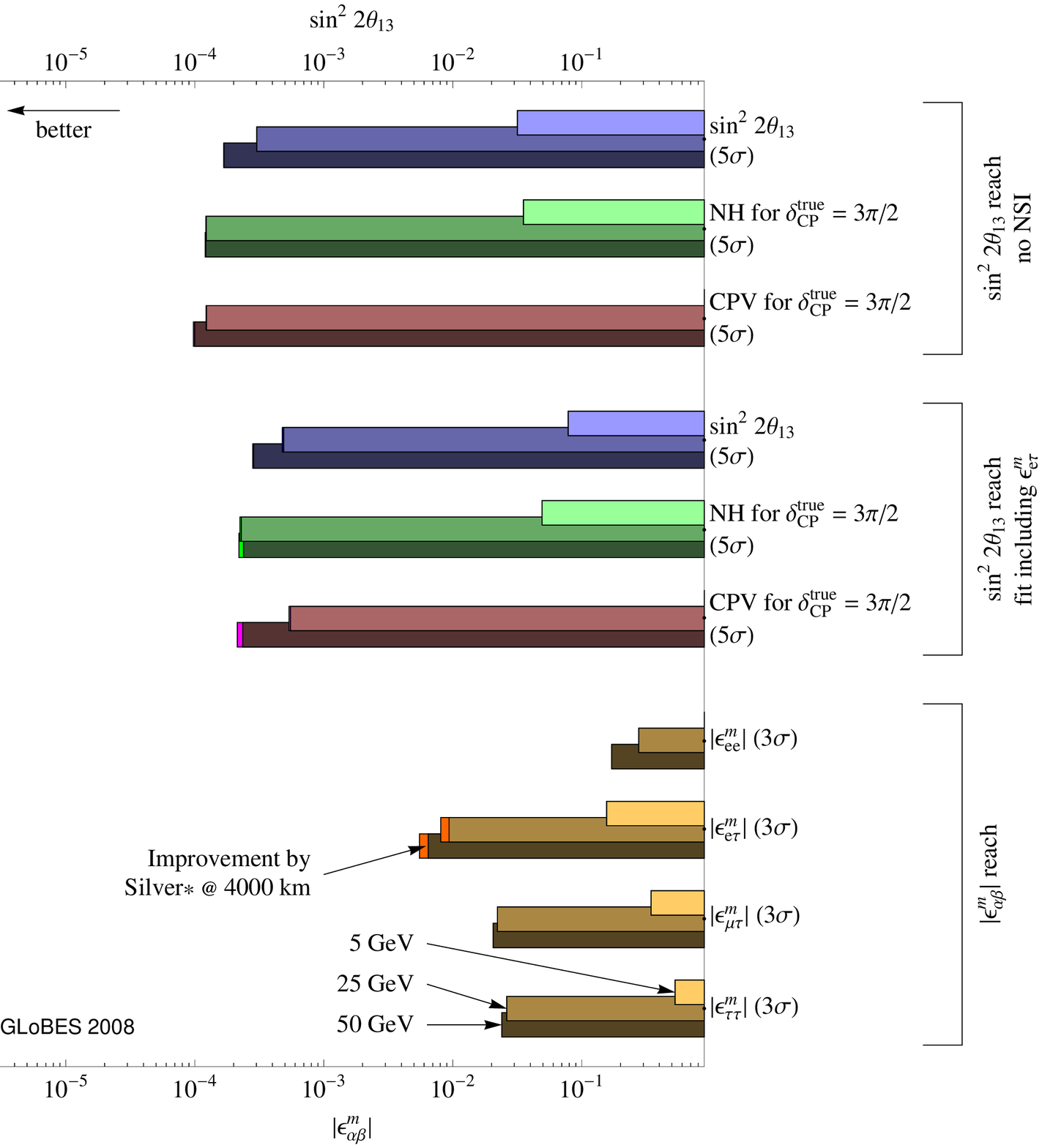}
  \end{center}
  \mycaption{\label{fig:summary}Summary for the optimization of a neutrino
    factory as a function of the muon energy. The dark bars represent $E_\mu=50 \, \mathrm{GeV}$, the medium light bars $E_\mu=25 \, \mathrm{GeV}$, and the light bars $E_\mu=5 \, \mathrm{GeV}$.  The upper group
of bars represents the standard optimization (in terms of the $\stheta$ reach), the middle group of bars represents the standard optimization (in terms of the $\stheta$ reach) including $\eet$ marginalized over, and the lower group the non-standard optimization (in terms of the $| \epsilon_{\alpha \beta}^m|$ sensitivity).
Here the {\sf IDS-NF} setup is used with two baselines at $4 \, 000 \, \mathrm{km}$ and $7 \, 500 \, \mathrm{km}$. Both the sensitivities without silver channel, as well as with an advanced silver channel detector {\sf Silver*} are shown in all cases. As a benchmark point, $\sthetat=0.001$ and $\deltacpt=3 \pi/2$ has been chosen, as well as a true normal hierarchy.
}
\end{figure}

\begin{table}
  \centering
  \begin{tabular}{lrrr}
    \hline
    Performance indicator & 90\% C.L. & $3\sigma$ C.L. & $5\sigma$ C.L. \\
    
    \hline
    \multicolumn{4}{l}{\bf Standard oscillation physics} \\
    $\stheta$                                             & $4.25\cdot 10^{-5}$ & $1.22\cdot 10^{-4}$ & $3.03\cdot 10^{-4}$ \\
    Normal hierarchy (for $\deltacpt = 3 \pi/2$) & $2.27\cdot 10^{-5}$ & $5.93\cdot 10^{-5}$ & $1.23\cdot 10^{-4}$ \\
    Maximal CPV ($\deltacpt = 3 \pi/2$, NH)     & $1.49\cdot 10^{-5}$ & $4.68\cdot 10^{-5}$ & $1.23\cdot 10^{-4}$ \\

    \hline
    \multicolumn{4}{l}{\bf Standard oscillation physics polluted
                       by non-standard $\boldsymbol{\eet}$} \\
    $\stheta$                                             & $8.13\cdot 10^{-5}$ & $2.04\cdot 10^{-4}$ & $4.88\cdot 10^{-4}$ \\
    Normal hierarchy (for $\deltacpt = 3 \pi/2$) & $4.00\cdot 10^{-5}$ & $1.01\cdot 10^{-4}$ & $2.29\cdot 10^{-4}$ \\
     Maximal CPV ($\deltacpt = 3 \pi/2$, NH)      & $4.69\cdot 10^{-5}$ & $1.39\cdot 10^{-4}$ & $5.52\cdot 10^{-4}$ \\

    \hline
    \multicolumn{4}{l}{\bf Non-standard oscillation physics
                       with real $\boldsymbol{\epsilon^m_{\alpha\beta}}$} \\
    $\eem$
         (with $\text{Im} \, \epsilon^m_{e\mu} = 0$)           & $[-1.77\cdot 10^{-3} ,$  &
                                                                 $[-3.70\cdot 10^{-3} ,$  &
                                                                 $[-6.33\cdot 10^{-3} ,$ \\
                                                               & $ 1.71\cdot 10^{-3}]$  &
                                                                 $ 3.26\cdot 10^{-3}]$  &
                                                                 $ 5.98\cdot 10^{-3}]$ \\
    $\eet$                                                                                                                  
         (with $\text{Im} \, \epsilon^m_{e\tau} = 0$)          & $[-4.46\cdot 10^{-3} ,$  &
                                                                 $[-9.18\cdot 10^{-3} ,$  &
                                                                 $[-1.37\cdot 10^{-2} ,$ \\
                                                               & $ 3.51\cdot 10^{-3}]$  &
                                                                 $ 5.98\cdot 10^{-3}]$  &
                                                                 $ 0.93\cdot 10^{-2}]$ \\
    $\emt$                                                                                                                
         (with $\text{Im} \, \epsilon^m_{\mu\tau} = 0$)        & $[-3.69\cdot 10^{-4} , $  &
                                                                 $[-6.76\cdot 10^{-4} , $  &
                                                                 $[-1.16\cdot 10^{-3} , $ \\ 
								& $ 3.68\cdot 10^{-4}]$  &
                                                                 $ 6.74\cdot 10^{-4}]$  &
                                                                 $ 1.15\cdot 10^{-3}]$ \\
    $\eee$                                   & $[-1.37\cdot 10^{-1},$ & $[-2.77\cdot 10^{-1},$ & $[-3.48\cdot 10^{-1},$ \\
                                             & $1.23\cdot 10^{-1}]$ & $2.26\cdot 10^{-1}]$ & $3.86\cdot 10^{-1}]$ \\
    $\emm$                                   & $[-1.90\cdot 10^{-2} ,$  &
                                                                 $[-2.64\cdot 10^{-2} , $  &
                                                                 $[-3.58\cdot 10^{-2} , $ \\
					      & $1.89\cdot 10^{-2}]$  &
                                                                 $2.59\cdot 10^{-2}]$  &
                                                                 $3.55\cdot 10^{-2}]$ \\
    $\ett$                                 & $[-1.90\cdot 10^{-2} ,$  &
                                                                 $[-2.62\cdot 10^{-2} ,$  &
                                                                 $[-3.57\cdot 10^{-2} ,$ \\
					   & $1.90\cdot 10^{-2}]$  &
                                                                 $ 2.62\cdot 10^{-2}]$  &
                                                                 $ 3.57\cdot 10^{-2}]$ \\

    \hline
    \multicolumn{4}{l}{\bf Non-standard oscillation physics
                       with complex $\boldsymbol{\epsilon^m_{\alpha\beta}}$} \\
    $|\epsilon^m_{e\mu}|$                                      & $3.41\cdot 10^{-3}$ & $5.71\cdot 10^{-3}$ & $8.08\cdot 10^{-3}$ \\
    $|\epsilon^m_{e\tau}|$                                     & $4.74\cdot 10^{-3}$ & $9.36\cdot 10^{-3}$ & $1.75\cdot 10^{-2}$ \\
    $|\epsilon^m_{\mu\tau}|$                                   & $1.80\cdot 10^{-2}$ & $2.22\cdot 10^{-2}$ & $3.31\cdot 10^{-2}$ \\
    \hline
  \end{tabular}
  \mycaption{\label{tab:summary}Summary of the sensitivities achievable in the IDS
    baseline setup ($L_1 = 4000$~km, $L_2 = 7500$~km, $E_\mu = 25$~GeV)
    without a silver channel detector for different CL (1 d.o.f.). 
The first two groups show the $\stheta$ reaches, the last two groups the $\epsilon_{\alpha \beta}^m$ reaches.
In the case of real $\epsilon^m_{\alpha \beta}$, we give the positive and negative limits separately.
Here $\sthetat=0.001$ and $\deltacpt=3 \pi/2$, as well as a normal true hierarchy have been used as a benchmark point.
Note that especially the $\eee$ sensitivity depends on this benchmark point. 
}
\end{table}

Our main results are summarized in \figu{summary}. Obviously, $E_\mu = 25$~GeV provides
    an excellent sensitivity to all standard and non-standard
    performance indicators, while lower energies (represented by the light bars) are unfavorable for {\em both}
standard and non-standard performance indicators.
 Higher muon energies or the addition of a silver channel detector
    at 4000~km do not yield any significant improvement, neither for standard oscillations, nor for NSI.
Furthermore, we show in \Tab~\ref{tab:summary} the expected sensitivities for the {\sf IDS-NF} baseline setup~1.0, \ie, $E_\mu=25 \, \mathrm{GeV}$, $L_1=4 \, 000 \, \mathrm{km}$, $L_2=7 \, 500 \, \mathrm{km}$, but without the silver channel.
For the standard oscillation parameters, the sensitivities become somewhat worse if there is an NSI pollution from $\eet$ (second group versus first group of sensitivities), but the orders of magnitude do not change.  In addition, we read off the table that for $\eema$ and $\emma$, the neutrino factory will hardly improve the current limits (see \Ref~\cite{Bandyopadhyay:2007kx} and references therein). For $\eeta$, the current limit can be improved by about two orders of magnitude, for $\emt$, about one order of magnitude (if $\emtp$ is assumed to be free), and for $\etta$, about two orders of magnitude.  For $\eee$, the improvement depends very much on  $\stheta$ and $\deltacpt$, as well as the matter density uncertainty, as we have discussed in \Sec~\ref{sec:muonen}. In general, a factor of a few may be expected. We believe that this overall performance is very impressive, but remember that the current bounds might be improved by the time a neutrino factory is actually built.

Of course, our study has been based on a particular type of new physics, namely 
non-standard neutral current interactions affecting the neutrino propagation.
One may ask the question whether, within the {\sf IDS-NF} baseline setup, our results
can be qualitatively generalized to other types of new physics.
In particular, are there scenarios which constitute a physics case for
$\nu_\tau$ detection?
In general, there might be two qualitatively different approaches
to search for new physics using the information from $\nu_\tau$
events:
\begin{enumerate}
\item
 Use the spectral dependence of the $\nu_\tau$ events and test its
 consistency with standard oscillations, and with new physics
 scenarios. This is what we have done in this study for the
 specific case of non-standard interactions in the $\nu_e
 \rightarrow \nu_\tau$ (silver) channel.
\item
 Test unitarity by using information from all flavors.
\end{enumerate}
For approach~1, the $\nu_\tau$ appearance channels can only
contribute significantly in the unlikely case that the impact
of the new physics is much larger in these channels than in
the others. After all, the golden and disappearance channels
provide much larger statistics, and will therefore typically
dominate the measurement, if appropriate baselines and
neutrino energies are used. Note that the $\nu_\mu
\rightarrow \nu_\tau$ channel may not be feasible at all
because the high event rates in this channel might prohibit
successful reconstruction.
For approach~2, one could either use neutral currents
as a signal (which would not require a dedicated $\nu_\tau$
detector), or consider the flavor sum of charged current event
rates. For the neutral currents, systematical uncertainties and
the charged current contamination will limit the measurement
to a precision of, perhaps, a few percent~\cite{Barger:2004db}.
For the charged currents at the neutrino factory, the weakest
link will probably be the detection of electron neutrinos
(preferably with charge identification), not the detection of
$\nu_\tau$. Electron neutrino events are very difficult to
reconstruct using an iron calorimeter, because electrons
produce electromagnetic showers. The associated uncertainty
in the event rates is expected to be of the order of a few
per cent, so that summing the charged current events over
all flavors will not yield a sensitivity significantly
exceeding that from neutral currents.
Therefore, we expect that the conclusions from this study
concerning the silver channel are likely to be translated
to many other new physics cases as well, which, however,
needs to be proven in specific studies. 

\subsection*{Acknowledgments}

JK would like to thank the Studienstiftung des deutschen Volkes for support.
TO and WW would like to acknowledge support from Emmy Noether program of Deutsche Forschungsgemeinschaft.

\end{document}